\pdfoutput=1

\documentclass[11pt,a4paper]{article}
\usepackage{jheppub}
\usepackage{amsmath}
\usepackage{mathrsfs}
\usepackage[utf8]{inputenc}
\usepackage{tikz}
\usepackage{tikz-3dplot}
\usepackage{multicol,bbm}
\usepackage{enumerate}
\usepackage{bibentry}
\usepackage{epstopdf}
\numberwithin{equation}{section}
\usepackage{graphicx}
\usepackage{subfigure}
\usepackage{overpic}

\usepackage{amssymb}
\usepackage{bm}
\usepackage{multirow}
\usepackage{caption}
\usepackage{tabularx}

\def\be{\begin{equation}}
\def\ee{\end{equation}}
\def\ba{\begin{eqnarray}}
\def\ea{\end{eqnarray}}

\def\CP1{\mathbb{CP}^1}
\def\SL2C{\mathrm{SL}(2,\mathbb{C})}

\def\Z2{\mathbb{Z}_2}

\def\su2{{SU(2)}}

\def\[{\left[}
\def\]{\right]}

\def\({\left(}
\def\){\right)}
\def\[{\left[}
\def\]{\right]}

\def\<{\langle}
\def\>{\rangle}

\def\i2{\frac{i}{2}}

\def\2F1{\,_2{\rm F}_1}

\begin{document}


\title{An Etude on Recursion Relations
and Triangulations 
}


\author[a,b]{Song He}\qquad
\author[c,a]{Qinglin Yang}
\affiliation[a]{CAS Key Laboratory of Theoretical Physics, Institute of Theoretical Physics, Chinese Academy of Sciences, Beijing 100190, China}
\affiliation[b]{School of Physical Sciences, University of Chinese Academy of Sciences, No.19A Yuquan Road, Beijing 100049, China}
\affiliation[c]{School of the Gifted Young, University of Science and Technology of China, No.96 Jinzhai Road, Hefei 230026, China}
\emailAdd{songhe@itp.ac.cn}
\emailAdd{yqlg3@mail.ustc.edu.cn}

\date{\today}

\abstract
{Following~\cite{Arkani-Hamed:2017thz}, we derive a recursion relation by applying a one-parameter deformation of kinematic variables for tree-level scattering amplitudes in bi-adjoint $\phi^3$ theory. The recursion relies on properties of the amplitude that can be made manifest in the underlying kinematic associahedron, and it provides triangulations for the latter. Furthermore, we solve the recursion relation and present all-multiplicity results for the amplitude: by reformulating the associahedron in terms of its vertices, it is given explicitly as a sum of ``volume" of simplicies for any triangulation, which is an analogy of BCFW representation/triangulation of amplituhedron for ${\cal N}=4$ SYM.}

\maketitle
\section{Introduction}
In~\cite{Arkani-Hamed:2017thz}, it has been shown that tree-level scattering amplitudes in various massless theories can be encoded in differential forms on the kinematic space, which is the space of Mandelstam variables in general spacetime dimension. In particular, amplitudes for bi-adjoint $\phi^3$ theory 
\cite{Cachazo:2013iea} are given by the canonical form/function~\cite{Arkani-Hamed:2017vfh} of an associahedron polytope~\cite{Stasheff_1,Stasheff_2} defined directly in kinematic space (equivalently the canonical function is the volume of the dual polytope). This gives a purely geometrical definition of bi-adjoint scalar amplitudes, in analogy with the amplituhedron of ${\cal N}=4$ SYM~\cite{Arkani-Hamed:2013jha, Arkani-Hamed:2017tmz}. Such differential forms can be constructed for tree amplitudes in any massless theories with color, which are linear combination of bi-adjoint $d\log$ forms with kinematic numerators as coefficients and naturally arise from worldsheet picture via scattering equations~\cite{Cachazo:2013gna, Cachazo:2013hca}.

In this new picture, the usual Feynman-diagram expansion in terms of cubic tree graphs corresponds to a particular way for triangulating the dual associahedron: each Feynman diagram is given by the ``volume" of a dual simplex associated with a facet of the dual associahedron (or a vertex of the associahedron itself). A generic triangulation of the dual associahedron also gives a representation of the amplitude with local poles only, and the Feynman-diagram expansion is the special one by introducing a point at infinity.  More interestingly, any triangulation of the associahedron itself produces totally new representations with spurious poles~\cite{Arkani-Hamed:2017thz}, in the same way as BCFW representations for gluon amplitudes or supersymmetric ones in {\it e.g.} ${\cal N}=4$ SYM \cite{BCF,BCFW}; for the latter, each BCFW term is given by one cell in a triangulation of the amplituhedron.
Due to the fact that each facet of associahedron is the product of two lower-point ones, the triangulation is done recursively, which in turn offers a geometric recursion relation that expresses bi-adjoint $\phi^3$ amplitude in terms of lower-point ones~\cite{Arkani-Hamed:2017thz}.

However, the analogy with the tree-level amplituhedron and BCFW representation in ${\cal N}=4$ SYM is not perfect yet for the following reasons. First, beyond $n=5$, no explicit result for the triangulation, {\it i.e.} solution to the geometric recursion, has been worked out in~\cite{Arkani-Hamed:2017thz}. This is in contrast with all-multiplicity solutions of BCFW recursion~\cite{DrummondHenn}, and the interpretation that for any $n$ each BCFW term can be identified with a simplex in the triangulation of the amplituhedron~\cite{Hodges, Arkani-Hamed2012}. Moreover, unlike the case of ${\cal N}=4$ SYM, no field-theoretical derivation based on an analogy of ``BCFW shift" is available for this geometry-motivated recursion for bi-adjoint $\phi^3$ amplitudes~\footnote{Of course, there exist other ``BCFW-like shifts" and recursion relations for scalar theories, such as those in~\cite{Cachazo:2016njl} and~\cite{Cheung:2015cba}, which have been applied to non-linear sigma model {\it etc.}. However, such recursion relations do not seem to apply to bi-adjoint $\phi^3$ amplitudes, neither are they related to the geometry. On the other hand, there are Berends-Giele recursion for $\phi^3$ amplitudes proposed in~\cite{Mafra:2016ltu}, but it is not of BCFW type.}. The geometric recursion for the canonical form of associahedra found in~\cite{Arkani-Hamed:2017thz} is the only known recursion that is applicable to something as simple as $\phi^3$ amplitudes!

It is thus an important open question how to derive such a recursion relation for cubic trees directly (without resorting to the geometry), and we fill the gap in this paper. In section~\ref{sec 2}, we provide a purely field-theoretical derivation of a recursion relation for $\phi^3$ amplitudes from a one-parameter deformation in kinematic variables, $X_{i j} \to z X_{i j}$ for $n{-}3$ independent (planar) Mandelstam variables. Remarkably one can then write a contour integral for the amplitude, as a meromorphic function of $z$, which has no pole at $z=0$ or $z \to \infty$. This is guaranteed by a special property of the $\phi^3$ amplitude, followed from the ``projectivity" of the canonical form~\cite{Arkani-Hamed:2017thz}; as a result, one can directly write the original amplitude, {\it i.e.} the residue at $z=1$, as a sum of residues at other $z$, which are products of lower-point amplitudes, thus providing the recursion relation. At every step, any choice of the variables to deform gives a different recursion, and each way of recursing down to lowest amplitudes gives a triangulation of the associahedron; as a side remark we also clarify when the corresponding triangulation is ``inside" and when one is ``outside". It is straightforward to solve the recursion, as we show in our explicit examples for $n=4,5,6$.

Explicit solutions to the recursion become more and more complicated as $n$ grows. However, in section~\ref{sec 3}, we present a compact formula for the amplitude for any $n$ as a sum of canonical functions of the simplices for any triangulation. After deriving a simple method for the coordinate of all {\it vertices} of the associahedron, we use it to compute the canonical function of any simplex in the associahedron. A sum of these simplices in any triangulation then gives the amplitude in ``BCFW representation" for bi-adjoint $\phi^3$ amplitude, which is very similar to that for ${\cal N}=4$ SYM tree as a sum over cells of the amplituhedron. As we show explicitly, each term there is like an ``R-invariant" of the NMHV tree amplitude, which contains at least one physical pole or external facet, and the remaining poles correspond to spurious ones or internal facets. Properties of the amplitude obscured by Feynman diagrams, such as the ``projectivity", become manifest term by term in BCFW representation. We present a general formula for any triangulation/solution to the recursion, including explicit examples up to $n=8$, and discuss general forms of spurious poles. We end the paper with discussions in section~\ref{sec 4}, and in the appendix we include results on triangulations of the dual associahedron and similar recursion/triangulations for Cayley polytopes which generalize the associahedron naturally~\cite{Gao:2017iop,He:2018pue}

\subsection{Review and notations}
Let's first recall the associahedron in kinematic space and its relation to bi-adjoint $\phi^3$ amplitudes~\cite{Arkani-Hamed:2017thz}. For space-time dimension $D\geq n{-}1$, the dimension of the kinematic space is $n(n{-}3)/2$, and it can be spanned by the so-called {\it planar variables} $X_{i j}$ for $1\leq i<j{-}1<n$. These variables are identified with the diagonals of an $n$-gon with edges given by $p_1, p_2, \cdots, p_n$, $X_{i j}=(\sum_{a=i}^{j{-}1} p_a)^2:=s_{i,i{+}1, \cdots, j{-}1}$. Each {\it planar} cubic tree graph corresponds to a full triangulation of the $n$-gon, with $n{-}3$ {\it compatible} $X_{i j}$'s being the inverse propagators of the graph. The {\it kinematic associahedron}, $\mathscr{A}(1,...,n):=\mathscr{A}_{n-3}$, is a $(n{-}3)$-dim polytope defined as the intersection of the positive region:
\begin{equation}
\Delta_n=\{X_{ij}\geq0\ {\rm for\ all}\ 1\leq i<j{-}1< n\}
\end{equation}
with the $(n{-}3)$-dim hyperplane defined by the following $(n{-}2)(n{-}3)/2$ conditions:
\begin{equation}
\begin{split}
H(1,2,\cdots,n):=\{C_{ij}=X_{ij}+X_{i+1\ j+1}-X_{i\ j+1}-X_{i+1\ j}\\
{\rm are\ positive\ constants, \, for}\ 1\leq i<j{-}1<n{-}1\}
\end{split}
\end{equation}
where we have excluded $C_{ij}$'s with $j=n$, but the resulting associahedron $\mathscr{A}_{n-3}=\Delta_n \cap H(1,2,\cdots,n)$ is actually cyclic invariant~\cite{Arkani-Hamed:2017thz}. Its canonical form is given by
\be
\Omega_{\mathscr{A}_{n-3}}=\Omega^{(n{-}3)}_{\phi^3}(1,2,\cdots,n)|_{H(1,2,\cdots,n)}=\prod_{a=1}^{n{-}3} d X_{i_a, j_a}~A(1,2,\cdots,n)\,,
\ee
where we use $A(1,2,\cdots,n):=m(1,2,\cdots,n |1,2,\cdots,n)$ to denote the ``diagonal" bi-adjoint amplitudes, given by the sum of all planar cubic trees; $\Omega_{\phi^3}(1,2,\cdots,n)$ is the {\it planar scattering form}~\cite{Arkani-Hamed:2017thz}, given by the sum of wedge products of $d\log$'s of $X$'s for planar cubic trees with signs to ensure that it is well-defined in a projectivized space ({\it i.e.} it only depends on ratio of $X$'s). For example for $\mathscr{A}_1$ is a line interval, and $\Omega_{\phi^3}(1,2,3,4)=d\log \frac{X_{13}}{X_{24}}$ (recall $X_{13}=s$, $X_{24}=t$);
by the pullback to $H(1,2,3,4)$ defined by $X_{13}+X_{24}=C_{13}$, we have
\be
\Omega_{\mathscr{A}_1}=d X_{13}~A(1,2,3,4)\,, \quad A(1,2,3,4)=\frac{1}{X_{13}}+ \frac{1}{C_{13}-X_{13}}=\frac 1 s +\frac 1 t\,.
\ee
Similarly $\mathscr{A}_2$ is a pentagon and the pullback of $\Omega_{\phi^3}(1,2,3,4,5)$ gives $A(1,2,3,4,5)$, which is the sum of five planar trees (see below). More generally $m(\alpha|\beta)$ can be given by the pullback of $\Omega_{\phi^3}(\alpha)$ to $H(\beta)$ for any pair of orderings $\alpha$ and $\beta$. Our discussion of the recursion relation and the solutions will mostly focus on $A(1,2,\cdots,n )$, but as we will show in the end they can be generalized to $m(\alpha|\beta)$ as well.

Before we derive recursion relations for $A_n:=A(1,2,\cdots,n)$, we first introduce some notation. The hyperplane can be parametrized by any $(n{-}3)$ independent $X_{i j}$'s, which will be called ``basis $X$'s" and denoted as $X:=\{X_{i_1, j_1}, X_{i_2, j_2}, \cdots, X_{i_{n{-}3}, j_{n{-}3}}\}$; any planar variable $X_{a,b}$ is a linear combination of the basis X's and the constants. We introduce the notation for the linear combination
\be\label{0}
X_{a b}=X^0_{a, b}+c_{a,b}\,,
\ee
where $X^0_{a, b}$ denotes the linear combination of basis $X$'s and $c_{a, b}$ for that of constants; both of which of course depends on the basis choice, but we suppress the explicit reference to that choice. In this notation, the amplitude becomes a function of the basis $X$'s and the constants $C$, which we denote as $A_n(X, C)$.

\paragraph{Diamond diagram of planar variables} To fully illustrate our construction, here we record explicit formulas for such linear combinations, which were first worked out in~\cite{Arkani-Hamed:2017thz}. These relations are most conveniently derived from the so-called ``diamond diagram" .Generally, an $n$-point diamond diagram is a triangular diagram with $n-3$ rows of the form:
\begin{equation}
\begin{matrix}
\phantom{a}& \phantom{a}& \phantom{a}& X_{1\ n-1}& C_{1\ n-1}& X_{2n}& \phantom{a}& \phantom{a}\\
\phantom{a}& \phantom{a}& ...& ...& ...& ...& ...& \phantom{a}\\
\phantom{a}& X_{14}& C_{14}& X_{25}& C_{25}& ...& C_{n-4\ n-2}& X_{n-3\ n}& \phantom{a}\\
X_{13}& C_{13}& X_{24}& C_{24}& X_{35}& ...& X_{n-3\ n-1}& C_{n-3\ n-1}& X_{n-2\ n}
\end{matrix}
\end{equation}
where every $C$ can be expressed as a combination of $X$'s around it, $C_{ij}=X_{ij}+X_{i+1\ j+1}-X_{i\ j+1}-X_{i+1\ j}$, and this directly leads to more general identities:
\begin{equation}\label{7}
X_{ik}+X_{jl}-X_{jk}-X_{il}=\sum_{\substack{i\leq a<j\\k\leq b<l}}C_{ab}
\end{equation}
{\it i.e.} this combination of four $X$'s of a quadrilateral is given by the sum of all $C$'s inside it. From (\ref{7}) it is easy to write any $X_{a,b}$ in terms of linear combination basis $X$'s and the $C$'s. Note that one has to choose the basis $X$'s to be linearly independent, or geometrically they intersect at a point; one convenient choice is to take the basis $\{X_{2n}, X_{3n}, \cdots, X_{n-2\ n}\}$. For example, for $n=5$ it is easy to read from the diagram that
\begin{equation}\label{5pt0}
\begin{split}
&X_{13}=-X_{25}+C_{13}+C_{14}\\
&X_{14}=-X_{35}+C_{14}+C_{24}\\
&X_{24}=-X_{35}+X_{25}+C_{24}
\end{split}
\end{equation} and for $n=6$ one finds the following linear combinations from the diagram:
\begin{equation}\label{6pt0}
\begin{split}
&X_{15}=-X_{46}+(C_{15}+C_{25}+C_{35})\\
&X_{13}=-X_{26}+(C_{13}+C_{14}+C_{15})\\
&X_{14}=-X_{36}+(C_{14}+C_{15}+C_{24}+C_{25})\\
&X_{25}=-X_{46}+X_{26}+(C_{25}+C_{35})\\
&X_{35}=-X_{46}+X_{36}+C_{35}\\
&X_{24}=-X_{36}+X_{26}+(C_{24}+C_{25})
\end{split}
\end{equation}

\section{Recursion Relations for $\phi^3$ Amplitudes}\label{sec 2}

\subsection{Derivation of recursion relations}

The basic idea is to introduce a one-parameter deformation in the kinematic space. Already implicitly used in~\cite{Arkani-Hamed:2017thz}, the most natural choice is to rescale the $n{-}3$ basis $X$'s by
\be
X_{i_a, j_a} \to \hat{X}_{i_a, j_a}:=z X_{i_a, j_a}\,,\quad {\rm for}~a=1,2,\cdots, n{-}3\,,
\ee
and keep the constants $C_{ij}$ unchanged. With this deformation, the amplitude $A_n(z X, C)$ becomes a meromophic function of $z$ on the complex plane, where all the poles of $A_n$, {\it i.e.} the $X_{a,b}$'s, become linear functions of $z$, $\hat{X}_{a,b}=z X^0_{a,b}+c_{a,b}$. Following the same logic as the derivation of BCFW recursion, we consider the following contour integral
\be
A_n(X, C)=\oint_{|z-1|=\varepsilon} \frac{z^{n{-}3} d z}{z-1} A_n(z X, C)\,,
\ee
where the original, un-deformed amplitude is given by the residue at $z=1$, and now one can use Cauchy theorem to write it as (minus) the sum of all other residues
\be
A_n(X, C)=-\left({\rm Res}_{z=\infty} + \sum_ {\rm finite\ poles} {\rm Res}_{z=z^*}\right)
\frac{z^{n{-}3} d z}{z-1} A_n(z X, C)
\ee
where we denote any pole of $A_n(z X, C)$ at finite position as $z^*$, and so far we have not used any properties of the amplitude or the associahedron. Now the first fact about the amplitudes is that there is no pole at infinity:
\be
\lim_{z\to \infty} z^{n{-}3} A_n(z X, C)={\cal O}(\frac {1}{z})\,,
\ee
One can easily verify this by expanding $A_n (z X, C)$ in terms of $1/z$: since all propagators are  the leading term amounts to setting all the $C$'s to 0:

\be
\lim_{z\to \infty} z^{n{-}3} A_n(z X, C)=\lim_{z\to \infty} z^{n{-}3} A_n (z X, 0)
\ee
and the latter vanishes due to a simple property of the canonical form/function of the associahedron $\mathscr{A}_{n{-}3}$. For any fixed $i$, by setting $n{-}3$ $C_{ij}=0$, the canonical function vanishes since the geometry $\mathscr{A}_{n{-}3}$ degenerates (all facets $X=0$ passes through the origin). Even without resorting the geometry, one can prove this property by factorization: any residue of $A_n(X, C)$ vanishes since lower-point amplitudes with $C=0$ vanishes, and we know trivially $A_4 (X, 0)=\frac{1}{X_{13}}+\frac{1}{0-X_{13}}=0$.

If we only care about the amplitude but not the form, it is totally natural to put a general numerator $z^m$, and there is no pole at infinity for any $0\leq m \leq n{-}3$. However, there is another special position, $z=0$: for $m=n{-}3$ obviously we have no contribution from the pole $z=0$, but for $m<n{-}3$, we may have such a pole, whose physical interpretation is less clear: generically the pole at $z=0$ can be of higher order, but even for the simple pole case where the contour integral gives a similar recursion relation or $A_n(X,C)$, they do not correspond to triangulations of the associahedron! Thus, to avoid complications introduced by the residue at $z=0$, we stick to the ``geometric" choice with $m=n{-}3$, where neither $z=0$ nor $\infty$ contributes to the contour integral. Now we have
\begin{equation}\label{1}
A_n(X, C)=-\sum_{(a, b) \neq (i_1, j_1), \ldots, (i_k, j_k) } {\rm Res}_{z=z_{a b}} \frac{z^{n{-}3} d z}{z-1} A_n (z X, C)\,,
\end{equation}
where we sum over all poles determined by $\hat{X}_{a b}= z X^0_{a b}+ c_{a b}=0$, {\it i.e.} $z_{a b}:=-c_{a b}/X^0_{a b}$ for all $X_{a b}$'s different from the basis ones $X_{i_1, j_1}, \cdots, X_{i_{n{-}3}, j_{n{-}3}}$. Now the residue is easy to compute, recall that on any such physical pole, the amplitude factorizes:\begin{equation}
\lim_{X_{a b}\to 0}X_{a b}\cdot A_{1,2, \cdots, n}=A_{a, \cdots, b{-}1,I}\times A_{I, b,\cdots, a{-}1}\,,
\end{equation}
with $I$ the internal particle with on-shell momentum $p_I=\pm \sum_{i=a}^{b{-}1} p_i$ ($p_I^2=\hat{X}_{a,b}=0$), thus the residue is given by the product of two lower-point amplitudes :
\be
-{\rm Res}_{z=z_{a b}} \frac{z^{n{-}3} d z}{z-1} A_n (z X, C)=\frac{z_{a b}^{n{-}3}}{X^0_{a b} (1-z_{a b})} A_{a, \cdots, b{-}1, I} A_{I, b, \cdots, a{-}1}\,.
\ee
with both amplitudes evaluated at $z_{ab}$. By the identity $X^0_{a b} (1-z_{a b})=X_{a b}$, we arrive at
\begin{equation}\label{2}
\boxed{A_{1,2, \cdots ,n}(X, C)=\sum_{(a,b)\neq (i,j)'s} \frac{z_{a b}^{n-3}}{X_{a b}}A_{a, \cdots ,b{-}1,I}(z_{a b} X, C) \times A_{I, b, \cdots, a{-}1}(z_{a b} X, C)}
\end{equation}
where $z_{a b}=-c_{a b}/X^0_{ab}$ for all $X_{ab}$ different from the basis ones. For simplicity, we also denote each factorization term on the RHS as $\mathscr{A}_{a, \cdots, b{-}1, I} \times \mathscr{A}_{I, b, \cdots, a{-}1}$ (to signify the corresponding facet of $\mathscr{A}_n$), which as we will show shortly is given by in the canonical function of this term in the triangulation. To illustrate \eqref{2}, we spell out examples for $n=4, 5, 6$.
\subsection{Examples for $n=4,5,6$}
\paragraph{case $n=4$}
For $A_{1234}$ if we choose $X_{13}$ to be the basis, then $X_{24}=C_{1 3}-X_{1 3}$ and the only term, $\mathscr{A}_{12I} \times \mathscr{A}_{I34}$, is the residues at the pole determined by $\hat X_{24}:=C_{13}-z_{24} X_{13}=0$, {\it i.e.}
\begin{equation}
z_{24}=\frac{C_{13}}{X_{13}}\,.
\end{equation}
Since three-point amplitudes $A_3=\pm 1$, \eqref{2} gives the result for $A_{1,2,3,4}$ immediately:
\begin{equation}\label{4pt}
A_{1234}=\frac{z_{24}}{X_{24}}=\frac{C_{13}}{(C_{13}-X_{13})X_{13}}=\frac{-u}{s t}\,.
\end{equation}
Choosing $X_{24}$ as the basis gives the same result. Geometrically, $\mathscr{A}_{1234}$ is a line interval, thus its triangulation has only one term (as opposed to the two terms from Feynman diagrams). This result will be used as a basic building block in recursion for higher-point amplitudes.
\paragraph{case $n=5$} For $A_{12345}$, we will write down recursion for two different choices of basis. First we choose $\{X_{25}, X_{35}\}$ as basis, and we need to sum over residues at $\hat{X}_{13}=0$, $\hat{X}_{14}=0$, and $\hat{X}_{24}=0$. They correspond to the following factorizations:
\begin{equation}
\mathscr{A}_{12 I}\times\mathscr{A}_{I34 5}+\mathscr{A}_{123 I}\times\mathscr{A}_{I4 5}+\mathscr{A}_{23 I}\times\mathscr{A}_{1I4 5}
\end{equation}
Solving the conditions to get $z_{13}, z_{14}, z_{24}$, then by \eqref{2} and \eqref{4pt} we have:
\ba\label{3}
A_{12345}(X, C)=&\frac{z_{13}^2}{X_{13}}(\frac1{\hat X_{14}(z_{13})}+\frac1{\hat X_{35}(z_{13})})+\frac{z_{14}^2}{X_{14}}(\frac1{\hat X_{13}(z_{14})}+\frac1{\hat X_{24}(z_{14})})
+\frac{z_{24}^2}{X_{24}}(\frac1{\hat X_{14}(z_{24})}+\frac1{\hat X_{25}(z_{24})})\nonumber\\
=&\frac{(C_{13}+C_{14}) (C_{14}+C_{24})}{X_{35} X_{13} 
(C_{13} X_{35}+C_{14} (X_{35}-X_{25})-C_{24} X_{25})}+\frac{C_{14}C_{24}}{X_{25} X_{24} 
(C_{14} (X_{25}-X_{35})+C_{24} X_{25})}\nonumber\\
&+\frac{C_{13}(C_{14}+C_{24})^2}{X_{14}
(C_{14}(X_{35}-X_{25})-C_{24}X_{25})(C_{13} X_{35}+C_{14}(X_{35}-X_{25})-C_{24} X_{25})}
\ea
This answer appears to be lengthy but it actually has very simple structures if we introduce a better notation for the two quadratic, spurious poles. Recall our notation \eqref{0} that given a basis, any $X$ is a sum of the basis $X^0$ part, and the constant $c$ part, then we recognize that the spurious poles for $n=5$ (and more generally as we will see shortly for $n=6$) are always of the form
\begin{equation}\label{5}
X_{i j}^{0} c_{k l}-X_{k l}^0 c_{i j}:=Y^{i j}_{k l}\,,
\end{equation}
for some $i, j$ and $k, l$ (note that $Y^{i j}_{k l}=-Y^{k l}_{i j}$). For example, the spurious pole in the first term is nothing but $X_{13}^{0} c_{14}-X_{14}^0 c_{13}$ since $X_{13}^0=-X_{25}$, $c_{13}=C_{13}+C_{14}$ and $X_{14}^0=-X_{35}$, $c_{14}=C_{14}+C_{24}$ by \eqref{5pt0}. Of course the $X^0$'s and $c$'s all depend on our basis choice so here the three $Y$'s are defined with respect to the basis $\{X_{25}, X_{35}\}$. Equipped with this notation, the $5-$ point result can be put into this suggestive form:
\be\label{4.1}
A_{12345}(X,C)=\frac{(C_{13}+C_{14}) (C_{14}+C_{24})}{X_{13}X_{35}Y_{14}^{13}}+\frac{C_{13}(C_{14}+C_{24})^2}{X_{14}Y_{14}^{13}Y_{14}^{24}}+\frac{C_{14}C_{24}}{X_{24}X_{25}Y_{14}^{24}}
\ee
where one can easily show that the poles $Y_{14}^{13}$ in the first and second terms cancel each other, and $Y_{14}^{24}$ in the second and third terms cancel each other.

Similarly, we can choose another basis, such as $\{X_{13}, X_{14}\}$; here we have three factorization terms that read:
\begin{equation}
\mathscr{A}_{23 I}\times\mathscr{A}_{1I4 5}+\mathscr{A}_{234\bar I}\times\mathscr{A}_{1I 5}+\mathscr{A}_{34 I}\times\mathscr{A}_{12I 5}
\end{equation}
After a similar calculation, we find that the spurious poles are now $Y_{25}^{35}$ and $Y_{25}^{24}$, defined with respect to the basis $\{X_{13}, X_{14}\}$, and the amplitude is
\begin{equation}\label{4}
\begin{split}
A_{12345}(X,C)=\frac{(C_{13}+C_{14}) (C_{14}+C_{24})}{X_{13} X_{35}Y_{25}^{35}}+\frac{C_{24} (C_{13}+C_{14})^2}{X_{25}Y_{25}^{35} Y_{25}^{24}}+\frac{C_{13} C_{14}}{X_{14} X_{24} Y_{25}^{24}}
\end{split}
\end{equation}
It is straightforward to check that \eqref{4.1} or \eqref{4} gives the same result.
\paragraph{case $n=6$}
Before proceeding to the relation to triangulations, let's present the result for $A_{1,...,6}$. Again we choose the basis to be $\{X_{26}, X_{36}, X_{46}\}$ and apply the recursion relation once to get the following factorization terms
\begin{equation}
\begin{split}
\mathscr A_{12I}\times\mathscr A_{I3456}+\mathscr A_{123I}\times\mathscr A_{I456}\\
+\mathscr A_{1234I}\times\mathscr A_{I56}+\mathscr A_{23I}\times\mathscr A_{1I456}\\
+\mathscr A_{34I}\times\mathscr A_{12I56}+\mathscr A_{234I}\times\mathscr A_{1I56}
\end{split}
\end{equation}
where we need (shifted) $5-$point amplitudes, and we can choose to write them in any way we like, {\it e.g.} \eqref{4.1} or \eqref{4}, or even the Feynman-diagram expansion
\begin{equation}
\frac1{X_{13}X_{35}}+\frac1{X_{25}X_{35}}+\frac1{X_{25}X_{24}}+\frac1{X_{14}X_{24}}+\frac1{X_{13}X_{14}}
\end{equation}
Without making any choices, we can instead simply put all denominators together for these $5-$pt amplitudes, and a direct computation gives,
\be
\begin{split}\label{6}
A_{1,...,6}=\frac{N_{13}}{X_{13}X_{36}X_{46}Y^{13}_{35}Y^{13}_{15}Y^{13}_{14}}+\frac{N_{14}}{X_{14}X_{46}Y^{14}_{24}Y^{14}_{15}Y_{14}^{13}}\\
+\frac{N_{15}}{X_{15}Y^{15}_{35}Y_{15}^{13}Y_{15}^{14}Y^{15}_{24}Y^{15}_{25}}+\frac{N_{24}}{X_{24}X_{26}X_{46}Y^{24}_{25}Y_{24}^{15}Y_{24}^{14}}\\
+\frac{N_{25}}{X_{25}X_{26}Y^{25}_{35}Y_{25}^{15}Y_{25}^{24}}+\frac{N_{35}}{X_{35}X_{36}X_{26}Y_{35}^{13}Y_{35}^{15}Y_{35}^{25}}
\end{split}
\ee
where the spurious poles from the recursion are again nicely of the form $Y^{ij}_{kl}$ (for the basis $\{X_{26}, X_{36}, X_{46}\}$), but the explicit forms of the numerators are quite lengthy, which we put in the appendix. However, as we will show now, these terms are nothing but the ``volume" for a partial triangulation of the associahedron corresponding to this first step of the recursion. The result simplifies significantly if we use the recursion again for $5$pt amplitudes, which gives a full triangulation.

\subsection{From recursion to triangulations}
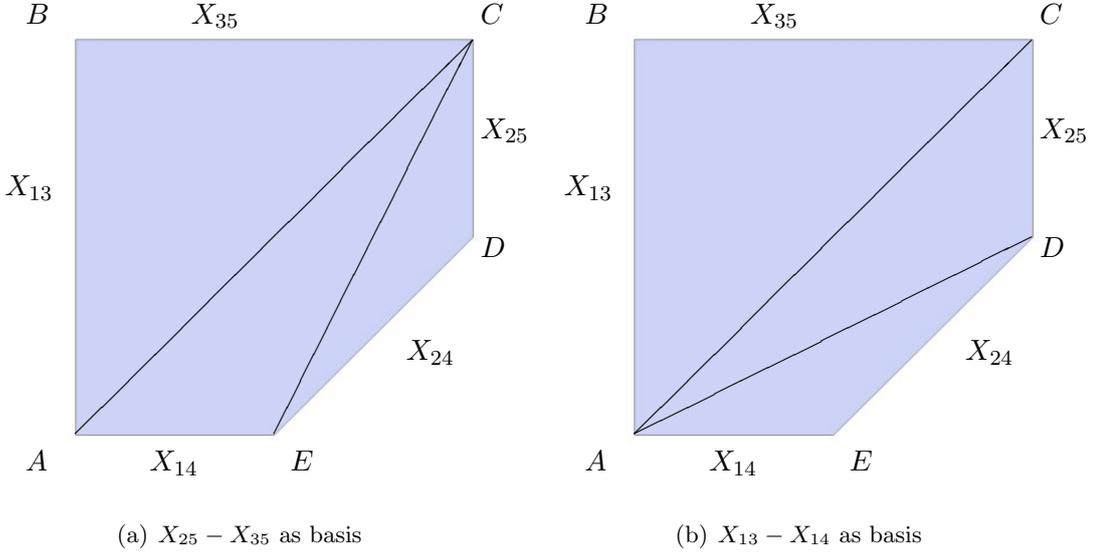
\begin{figure}[htbp]
\centering
\subfigure[$X_{25}-X_{35}$ as basis]{
\begin{minipage}{0.3\linewidth}
\centering
\begin{overpic}[width=1.2\textwidth]
{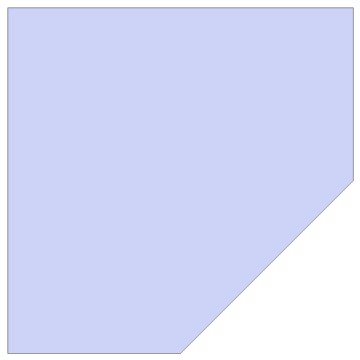}
\put(20,-7){$X_{14}$}
\put(-15,60){$X_{13}$}
\put(30,102){$X_{35}$}
\put(100,74){$X_{25}$}
\put(82,20){$X_{24}$}
\put(-10,-7){$A$}
\put(-10,102){$B$}
\put(100,102){$C$}
\put(100,45){$D$}
\put(54,-7){$E$}
\put(2,2){\color{black}\line(1,1){96}}
\put(50,2){\color{black}\line(1,2){48}}
\end{overpic}
\linebreak
\linebreak
\end{minipage}
}
\phantom{aaaaaaaaaaaa}
\subfigure[$X_{13}-X_{14}$ as basis]{
\begin{minipage}{0.3\linewidth}
\centering
\begin{overpic}[width=1.2\textwidth]
{5pt.jpg}
\put(20,-7){$X_{14}$}
\put(-15,60){$X_{13}$}
\put(30,102){$X_{35}$}
\put(100,74){$X_{25}$}
\put(82,20){$X_{24}$}
\put(-10,-7){$A$}
\put(-10,102){$B$}
\put(100,102){$C$}
\put(100,45){$D$}
\put(54,-7){$E$}
\put(2,2){\color{black}\line(1,1){96}}
\put(2,2){\color{black}\line(2,1){96}}
\end{overpic}
\linebreak
\linebreak
\end{minipage}
}
\caption{"inside" triangulations of $5$pt associahedra}\label{5pt instri}
\end{figure}

Before we turn to the discussion of solutions to the recursion in general, here we should first interpret our results for $n=5,6$ as triangulations of associahedra~\cite{Arkani-Hamed:2017thz}. 
Note that so far we have always chosen a basis where $X_{i_1, j_1}, \cdots, X_{i_{n{-}3}, j_{n{-}3}}$ are compatible planar variables, {\it i.e.} they are the poles of a planar cubic tree or diagonals of a triangulation of the $n$-gon. Geometrically, such a basis corresponds to a {\it vertex} of the associahedron, and the recursions, such as \eqref{4.1} (or \eqref{4}) and \eqref{6}, correspond to triangulations which uses this vertex as a reference point (or the``origin'). We call such triangulations ``inside" since everything is inside the convex associahedron polytope, and we will also discuss ``outside" triangulations, {\it i.e.} triangulations with non-compatible basis, shortly in the end.
\paragraph{``Inside" triangulations for $n=5,6$}
It is easy to see that our results for $n=5$, \eqref{4.1} and \eqref{4}, correspond to the two triangulations shown in figures \ref{5pt instri} using two different reference points, $\{X_{25}, X_{35}\}$ and $\{X_{13}, X_{14}\}$, respectively. Each term in the triangulation is the ``volume" of a triangle formed by the origin and a facet (here just an edge) that is not adjacent. The physical poles $X_{13}, X_{24}, \cdots, X_{25}$ are of course the five external facets (edges) of the pentagon, and the two spurious poles are the internal facets (diagonal lines) in each triangulation. In fact, each $Y^{i j}_{k l}$ corresponds the diagonal line connecting the origin to the vertex given by $\{X_{i j}, X_{k l}\}$.  Altogether there are 5 such ``inside" triangulations, each with a vertex as the origin.
\\

Similarly, the $6$pt result \eqref{6} represents a partial triangulation of the $n=6$ associahedron (Fig. \ref{cov 5,6} (b)) into 6 polytopes, where we connect the origin, $\{X_{26}, X_{36}, X_{46}\}$, to 6 facets that are not adjacent to it (those different from $X_{26}, X_{36}, X_{46}$). Note that 2 of them are quadrilaterals $\mathscr{A}_1 \times  \mathscr{A}_1$ and 4 are pentagons $\mathscr{A}_2 \times \mathscr{A}_0$. One can check that the 6 terms in \eqref{6} are given by the canonical functions of this 6 polytopes; as we will show shortly, it is much easier to further triangulate these pentagons and quadrilaterals to get full triangulations, which correspond to further use of recursion relations. In addition to the 6 external facets (local poles), we have 10 internal facets in this partial triangulation, in 1:1 correspondence with the 10 spurious poles  $Y^{i j}_{k l}$ in \eqref{6}; these internal ones are always triangles formed by connecting the origin to an edge of the associahedron, which is the intersection of $X_{ij}$ with $X_{k l}$.

Of course, we can choose any one of the 14 vertices as the origin, which gives a different partial triangulation of the associahedron. It is then an interesting combinatoric question to count the number of full ``inside" triangulations of $\mathscr{A}_3$. Note that there are two types of vertices: there are 12 vertices, each of which is not adjacent to 4 pentagons and 2 quadrilaterals (like the one we have chosen), and each of the remaining 2 vertices is not adjacent to 3 pentagons and 3 quadrilaterals; recall that we have 5 ways of further triangulating a pentagon, but only 2 ways of triangulating a quadrilateral, thus altogether we have
\begin{equation}
2\times2^3\times5^3+12\times2^2\times5^4=32000
\end{equation}
ways of full triangulations of the associahedron that are inside.

In general, we conclude that for any $n$, a choice of $X$ basis that are compatible with each other (thus the origin is one of the vertices) always give us an ``inside" triangulation:Each term on the RHS of (\ref{2}) is exactly the canonical function of a polytope obtained by connecting the origin to one of the facets that are not adjacent to it; Each spurious pole corresponds to a internal facet obtained by connecting the origin to a co-dimension 2 boundary (intersection of two external facets).

\begin{figure}[ht]
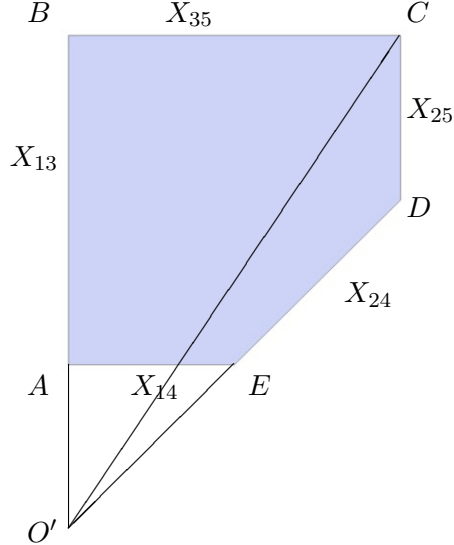

\centering
\begin{overpic}[width=0.3\textwidth]
{5pt.jpg}
\put(20,-7){$X_{14}$}
\put(-15,60){$X_{13}$}
\put(30,102){$X_{35}$}
\put(100,74){$X_{25}$}
\put(82,20){$X_{24}$}
\put(-10,-7){$A$}
\put(-10,102){$B$}
\put(100,102){$C$}
\put(100,45){$D$}
\put(54,-7){$E$}
\put(-10,-50){$O^\prime$}
\put(2,2){\color{black}\line(0,-1){48}}
\put(50,2){\color{black}\line(-1,-1){48}}
\put(2,-46){\color{black}\line(2,3){96}}
\end{overpic}
\linebreak
\linebreak
\linebreak
\linebreak
\caption{"outside" triangulation of $5$pt associahedron, $X_{13}-X_{24}$ as basis}
\label{outtri}
\end{figure}

\paragraph{``Outside" triangulations}
If we have chosen a basis of $X$'s that are not mutually compatible, the recursion leads to a different type of triangulation: the reference point or origin, which is the intersection of the $X$'s, lies outside the associahedron, thus it gives a triangulation that involve faces of all dimensions that are outside. For instance, choosing $\{X_{13}, X_{24}\}$ as a basis, the $5-$points amplitude becomes:
\begin{equation}
\begin{split}
A_{1,2,3,4,5}=\frac{C_{24} (C_{13}+C_{14})}{X_{24} X_{25} Y_{25}^{35}}+\frac{(C_{13}+C_{14}) (C_{13}+C_{14}+C_{24})}{X_{13}X_{35} Y_{25}^{35}}-\frac{C_{13}}{X_{13} X_{24} X_{14}}\\
=[Z_{O^\prime}Z_{B}Z_{C}]+[Z_{O^\prime}Z_{C}Z_{D}]-[Z_{O^\prime}Z_{A}Z_{E}]\phantom{aaaaaaaaaaaaaaaaaaaa}
\end{split}
\end{equation}
where $Z_O$ is the new origin, and we have introduced the notation $[Z_1 Z_2 Z_3]$ to denote (canonical function of) the triangle formed by connecting the three points $Z_1, Z_2, Z_3$ (see Fig. \ref{outtri}). We call such a triangulation an ``outside" one. 
\\

\section{Solutions of the Recursion and Triangulations}\label{sec 3}

We have seen that by choosing a $X$ basis to deform, applying the recursion \eqref{2} once leads to a new representation of $A_{1, 2, \cdots, n}$, which corresponds to a partial triangulation of the associahedron $\mathscr{A}_{n-3}$. On the other hand, though the explicit results for $n=4,5$ are simple enough, that for $n=6$ is already a bit complicated, thus it seems difficult to generalize to arbitrary $n$. However, we know that by repeatedly using the recursion, any ``BCFW" representation for the $\phi^3$ amplitude must corresponds to a full triangulation, where each term is given by the canonical function of a simplex; we denote the latter as a ``R-invariant" by the analogy with ${\cal N}=4$ SYM. In this section, we present the structure for a general triangulation as a solution to recursion for general $n$, where each ``R-invariant" can be worked out explicitly given our new formula for all the vertices of the associahedron.
\\

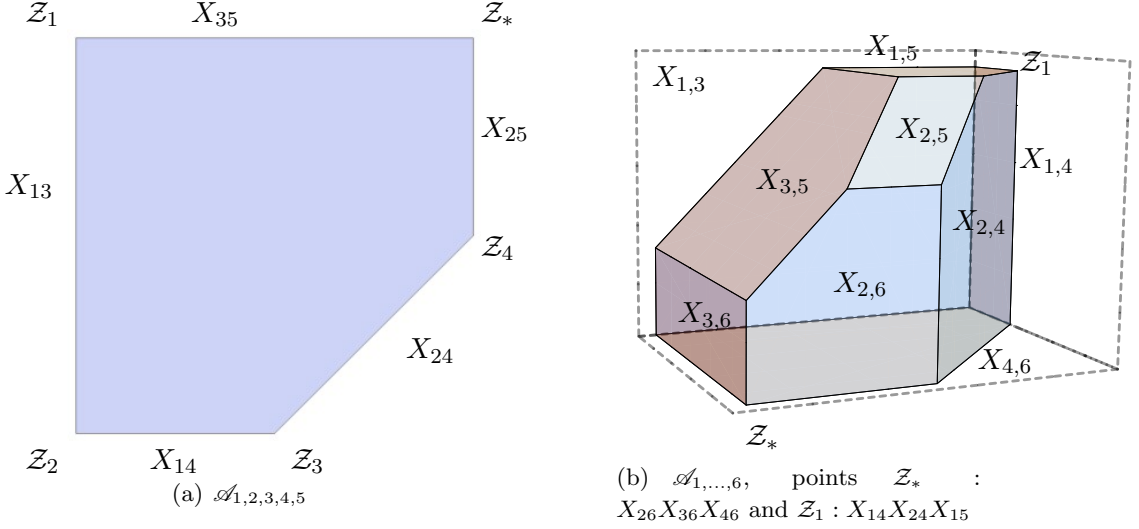
\begin{figure}[htbp]
\centering
\subfigure[$\mathscr{A}_{1,2,3,4,5}$]{
\begin{minipage}{0.3\linewidth}
\centering
\begin{overpic}[width=1.2\textwidth]
{5pt.jpg}
\put(20,-7){$X_{14}$}
\put(-15,60){$X_{13}$}
\put(30,102){$X_{35}$}
\put(100,74){$X_{25}$}
\put(82,20){$X_{24}$}
\put(-10,-7){${\cal Z}_2$}
\put(-10,102){${\cal Z}_1$}
\put(100,102){${\cal Z}_*$}
\put(100,45){${\cal Z}_4$}
\put(54,-7){${\cal Z}_3$}
\end{overpic}
\linebreak
\end{minipage}
}
\phantom{aaaaaaaaaaaa}
\subfigure[$\mathscr{A}_{1,...,6}$, points ${\cal Z}_*:X_{26}X_{36}X_{46}$ and ${\cal Z}_1:X_{14}X_{24}X_{15}$]{
\begin{minipage}{0.3\linewidth}
\centering
\begin{overpic}[width=1.5\textwidth]
{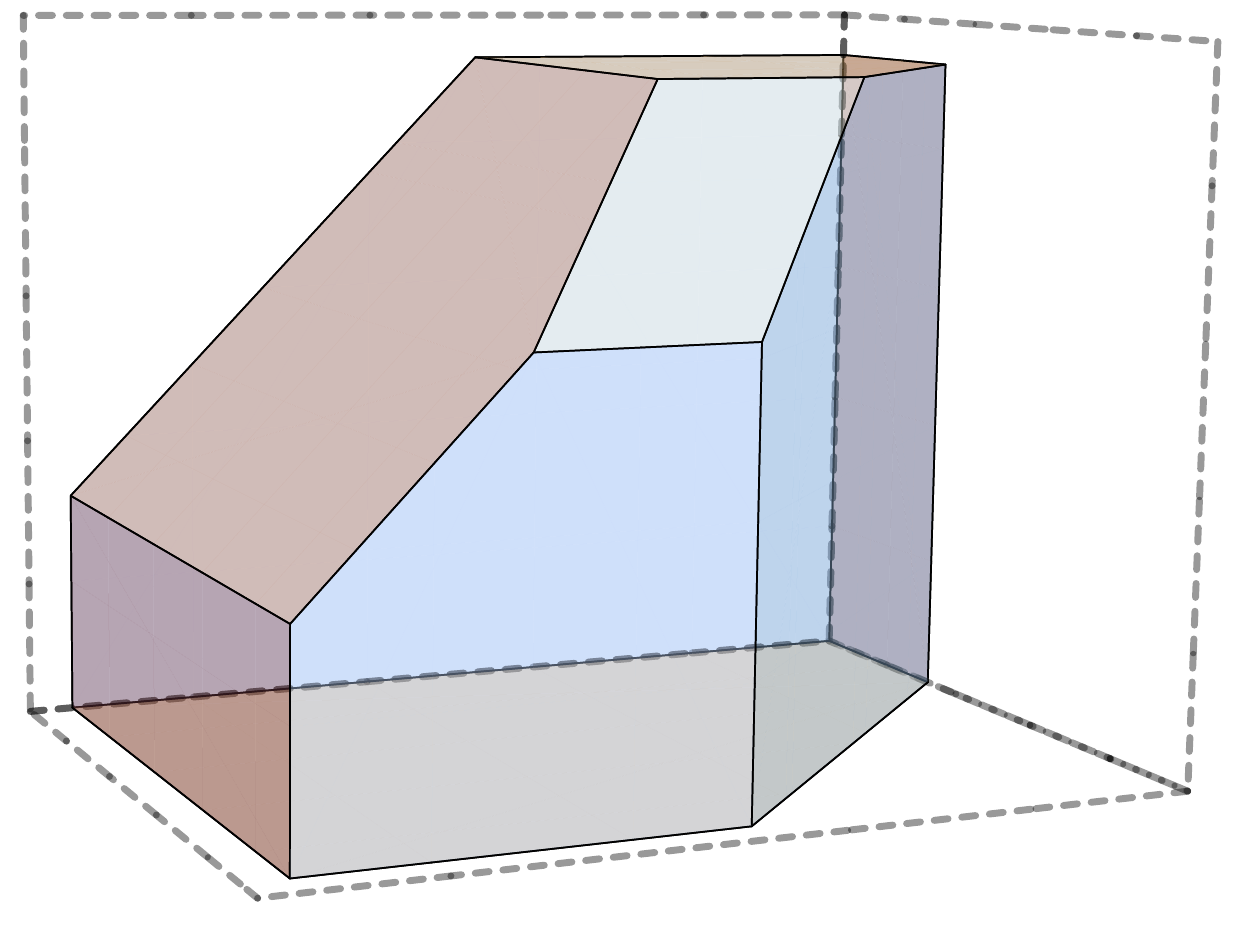}
  \put(5,66){$X_{1,3}$}
  \put(76,50){$X_{1,4}$}
  \put(68,11){$X_{4,6}$}
  \put(46,72){$X_{1,5}$}
  \put(25,46){$X_{3,5}$}
  \put(52,56){$X_{2,5}$}
  \put(40,26){$X_{2,6}$}
  \put(10,20){$X_{3,6}$}
  \put(63,38){$X_{2,4}$}
  \put(23,-4){${\cal Z}_*$}
  \put(76,69){${\cal Z}_1$}
\end{overpic}
\linebreak
\end{minipage}
}
\caption{$5,6$pt associahedra}\label{cov 5,6}
\end{figure}
\paragraph{The coordinate for vertices of the associahedron}
We have defined the kinematic associahedron $\mathscr{A}_{n{-}3}$ in terms of its facets $X_{a b}=X_{a b}^0+c_{a b}>0$. It is more convenient for the discussion of triangulations to define it in terms of vertices. Once a basis is chosen, we can obtain the coordinates of any vertex by solving for the $n{-}3$ basis $X_{i,j}$'s, $n{-}3$ linear equations of the form $X_{a,b}=0$; the solution expresses each basis element $X_{a,b}$ as a linear combination of the constants, and the latter is exactly the coordinate for this vertex,  which we denote as ${\cal Z}(\{X_{a,b}\})$. For example, with $X_{13}$ as basis and relation $X_{24}=C_{13}-X_{13}$, the coordinates of the two endpoints of $\mathscr{A}_1$ are ${\cal Z}(X_{13})=0$ and ${\cal Z}(X_{24})=C_{13}$. By definition, the coordinate for the reference point (origin) is $(0, \cdots, 0)$, which is also a vertex for the ``inside" version.

The first nontrivial case is $\mathscr{A}_2$ , and one can easily work out the coordinates, say, for the basis $\{X_{25}, X_{35}\}$. Let us denote the vertices $\{X_{25}, X_{35}\}$, $\{X_{13}, X_{35}\}$, $\{X_{13}, X_{14}\}$,  $\{X_{14}, X_{24}\}$, $\{X_{24}, X_{25}\}$ as ${\cal Z}_*, {\cal Z}_1, {\cal Z}_2, {\cal Z}_3, {\cal Z}_4$ respectively (see Fig. \ref{cov 5,6} (a)). For example, for vertex ${\cal Z}_2$ a direct computation by solving
\be
\begin{split}
&-X_{25}+C_{13}+C_{14}=0\\
&-X_{35}+C_{14}+C_{24}=0
\end{split}
\ee
gives ${\cal Z}_2=(C_{13}+C_{14},C_{14}+C_{24})$. Similarly we have ${\cal Z}_1=(C_{13}+C_{14},0)$, ${\cal Z}_*=(0,0)$, ${\cal Z}_4=(0, C_{24})$ and ${\cal Z}_3=(C_{14}, C_{14}+C_{24})$.
Moving to the next example for vertices of the $n=6$ case, $\mathscr{A}_3$ (see Fig. \ref{cov 5,6} (b)). With the basis $\{X_{26}, X_{36}, X_{46}\}$, the origin ${\cal Z}_*$ has coordinate $(0,0,0)$, and that for ${\cal Z}_1:=Z(\{X_{14}, X_{24}, X_{15}\}$ can be obtained by solving
\begin{equation}
\begin{aligned}
-X_{36}+(C_{14}+C_{15}+C_{24}+C_{25})&=0\\
-X_{36}+X_{26}+(C_{24}+C_{25})&=0\\
-X_{46}+(C_{15}+C_{25}+C_{35})&=0
\end{aligned}
\end{equation}
which gives \begin{equation}
{\cal Z}_1=(C_{14}+C_{15},C_{14}+C_{15}+C_{24}+C_{25},C_{15}+C_{25}+C_{35})\,.
\end{equation}
In appendix, we present a new mutation rule, which in principle gives a closed-formula for the coordinate (solution of the linear equations) for any vertex of the associahedron. Given the vertices, the associahedron can be defined as their convex hall: introducing affine coordinate, $Z:=(1, {\cal Z})$, then $\mathscr{A}_{n{-}3}$ is defined as the region constrained by
\be
Y:=(1, {\bf X})=\sum_{i=1}^{{\bf C}_{n{-}2}} \alpha_i~Z_i(\{C\})\,, \quad {\rm for}~\alpha_i>0,\quad \sum_i \alpha_i=1\,.
\ee
where ${\bf X}$ denotes the basis, and $Z_1(\{C\}), \cdots, Z_{{\bf C}_{n{-}2}}(\{C\})$ denotes the vertices above.
\subsection{General formula for amplitudes from triangulations}

Now we are ready to write down a general formula for {\it any} solution to the recursion relation, which represents the amplitude as a sum of canonical functions of simplices in a full triangulation of the associahedron. Let's first recall the definition of the canonical function for a simplex of dimension $n{-}3$. Denote the vertices of the simplex as $Z_0, Z_1, \cdots, Z_{n{-}3}$, its canonical function is defined as
\begin{equation}\label{10}
\boxed{[Z_{0},Z_{1},...,Z_{n{-}3}]=\frac{\langle Z_{0}Z_{1} \cdots Z_{n{-}3} \rangle ^{n{-}3}}{\prod_{i=0}^n \langle Y Z_{0}...\hat Z_{i}...Z_{n{-}3} \rangle}}
\end{equation}
where the bracket $\langle \cdots \rangle$ denotes the determinant of $n{-}2$ vectors of dimension $n{-}2$, and the hat denotes omission. Note that the canonical form is given by dressing it with $\langle Y d^{n{-}3} Y\rangle=d^{n{-}3} X$, which is a $d\log$ form of $n{-}3$ ratios of $n{-}2$ brackets in the denominator:
\be
\langle Y d^{n{-}3} Y\rangle[Z_0, Z_1, \cdots, Z_{n{-}3}] =\pm \bigwedge^{n{-}3} d\log\frac{<YZ_{0}...\hat Z_{i}...Z_{n}>}{<YZ_{0}...\hat Z_{j}...Z_{n}>}\,.
\ee
Given any triangulation of $\mathscr{A}_{n{-}3}$ into simplices which we label by $\Gamma$, we have
\be
A_{1,2,\cdots, n}=\sum_{\Gamma} {\rm sgn}_\Gamma [Z_0^\Gamma, Z_1^\Gamma, \cdots, Z_{n{-}3}^\Gamma]\,,
\ee
where the sign is determined by orientations of such simpleces, and we will see that for the ``inside" version they can be nicely chosen such that we have ${\rm sgn}_\Gamma=1$.

This is of course very familiar from the language of the amplituhedron in ${\cal N}=4$ SYM~\cite{Arkani-Hamed:2013jha} and more generally that of canonical forms of positive geometries~\cite{Arkani-Hamed:2017vfh}. The canonical function/form can be viewed as the "R-invariant" of the $\phi^3$ amplitudes, while the brackets are the analog of 4-brackets of momentum twistors. Recall that for NMHV tree amplitudes in ${\cal N}=4$ SYM, the amplituhedron is a four-dimensional (cyclic) polytope thus each R-invariant, or canonical form of a 4-simplex, involves 5 vertices. Here each R-invariant involves $n{-}2$ vertices since it is given by a $n{-}3$ dimensional simplex. The formula also makes it clear that any pole (physical or spurious) is linear in the basis $X$ (and polynomial in $C$'s), and the numerator of each simplex is a polynomial of $C$'s only.\\
Let's see the example for $n=5$, where we have a triangulation of pentagon:
\be\label{5pt}
A_{12345}=[Z_\star ,Z_1,Z_2]+[Z_\star, Z_2,Z_3]+[Z_\star ,Z_3,Z_4]\,,
\ee
and let's work out the canonical function of these triangles. For instance in the numerator of the first term $[Z_\star, Z_1 ,Z_2]$ of (\ref{5pt}), we have
\be
<Z_\star Z_{1}Z_{2}>=
\Biggm|\begin{matrix}
1 &0 &0\\
1 &C_{13}+C_{14} &0\\
1 &C_{13}+C_{14} &C_{14}+C_{24}\\
\end{matrix}\Biggm|
=(C_{13}+C_{14})(C_{14}+C_{24})
\ee
and its three denominators are
\be
<YZ_\star Z_{1}>=
\Biggm|\begin{matrix}
1 &X_{25} &X_{35}\\
1 &0 &0\\
1 &C_{13}+C_{14} &0\\
\end{matrix}\Biggm|
=(C_{13}+C_{14})X_{35}
\ee
and
\be
<YZ_\star Z_{2}>=(C_{13}+C_{14})X_{35}-(C_{14}+C_{24})X_{25}
\ee
\be
<YZ_1 Z_{2}>=(C_{14}+C_{24})X_{13}
\ee
which gives identical result as the first term of the recursion result \eqref{4.1}:
\be
[Z_\star, Z_1,Z_2]=\frac{<Z_\star Z_{1}Z_{2}>^2}{<YZ_\star Z_{1}><YZ_\star Z_{2}><YZ_1 Z_{2}>}=\frac{(C_{13}+C_{14}) (C_{14}+C_{24})}{X_{13}X_{35}Y_{14}^{13}}
\ee
Similarly we see that the other two terms of \eqref{4.1} are also nicely given by $[Z_\star ,Z_2 ,Z_3]$ and $[Z_\star ,Z_3 ,Z_4]$ respectively. This has essentially trivialized our previous calculation for solving the recursion for $n=5$.

In general, a triangulation of the associahedron can be achieved by performing $n{-}3$ steps of recursion as follows. In the first step we connect a origin, say $Z^{(0)}$, to all co-dimension-one facets that are not adjacent to $Z^{(0)}$, denoted as $X_{i_1, j_1}$, and sum over $(n{-}2)(n{-}3)/2$ such pairs $i_1, j_1$; in the second step, for each facet $X_{i_1, j_1}$, we triangulate it by connecting a origin $Z^{(1)}_{i_1,j_1}$ (the subscript is a reminder that the vertex belongs to this facet) to all of its (co-dimension-two) facets, denoted by $X_{i_2, j_2}$ (compatible with $X_{i_1, j_1}$), and sum over the pair $i_2, j_2$. Continuing this process till the last, $(n{-}3)$-th step, where we are left with a unique vertex $Z^{(n{-}3)}_{i_{n{-}3}, j_{n{-}3}}$, denoted by $X_{i_{n{-}3}, j_{n{-}3}}$ that is compatible with $X_{i_1, j_1}, \cdots, X_{i_{n{-}4}, j_{n{-}4}}$, and we have the amplitude (canonical function) as a nested sum:
\begin{equation}\label{8}
\boxed{A_{123...n}=\sum_{i_1, j_1} \sum_{i_2, j_2} \cdots \sum_{i_{n{-}4}, j_{n{-}4}}
[Z^{(0)}, Z^{(1)}_{i_1, j_1},\cdots,Z^{(n{-}4)}_{i_{n{-}4}, j_{n{-}4}}, Z^{(n{-}3)}_{i_{n{-}3}, j_{n{-}3}}]}
\end{equation}
where we have restricted to the ``inside" case with the origin at each step chosen to be a vertex, and it is straightforward to show that each ``R invariant" comes with a $+$ sign. For ``outside" triangulation, some of these terms will have $-$ sign, as we already discussed in the $n=5$ example. Another comment is that there is no sum over the last pair,  $i_{n{-}3}, j_{n{-}3}$ since there is no need to triangulate each edge; in fact, after the $n{-}5$-th step, we are left with two-dimensional faces (quadrilateral or pentagon), which are triangulated by (two or three) triangles thus for each such face, we can locally denote the last two $Z$'s as $Z^{(n{-}4)}=Z_a$, $Z^{(n{-}3)}=Z_{a{+}1}$ with denote the sum over $i_{n{-}4}, j_{n{-}4}$ as a sum over $a$ instead.

\subsection{Examples up to $n=8$}
Equation (\ref{8}) is a powerful formula from which we obtain explicit BCFW representation for bi-adjoint amplitudes to all multiplicities. We now take n=6,7,8 as examples to explain it in detail.
\paragraph{case $n=6$}Now we come back to the $n=6$ case. After we are equipped the equation \eqref{8}, $6-$pt amplitude can be directly written as:
\begin{equation}
A_{123456}(X,C)=\sum_{i_1, j_1} \sum_{a}[Z^{(0)}, Z^{(1)}_{i_1, j_1},Z_a,Z_{a+1}]
\end{equation}
Here, $Z^{(0)}$ is the origin points shared by the facets selected as basis in the first step of the recursion, {\it i.e.} $Z^{(0)}=(1,0,0,0)$. $ Z^{(1)}_{i_1, j_1}$  denotes the origin in the second step for facet $X_{i_1,j_1}$, where we sum over 6 pairs $i_1, j_1$ for $6$ facets that are not connected to $Z^{(0)}$. In the second sum, we can already sum over edges ($Z_aZ_{a+1}$) that are not connected to the point $Z^{(1)}_{i_1, j_1}$, for every face $X_{i_1, j_1}$ locally. In the end, $[Z_1,Z_2,Z_3,Z_4]$ stands for the canonical function of the simplex founded by the vertices $Z_i$, $i=1...4$.

Take $\mathscr{A}_{23I}\times\mathscr{A}_{I4561}$ as an example. By solving the coordinate of the vertices and summing over 3 "R-invariants", the amplitude for this term reads:
\begin{equation}\label{12}
[Z^{(0)}Z^{(1)}_{24}Z_{1}Z_{2}]+[Z^{(0)}Z^{(1)}_{24}Z_{2}Z_{3}]+[Z^{(0)}Z^{(1)}_{24}Z_{3}Z_{4}]
\end{equation}
where the terms denote the following 3 factorizations (3 triangles of the pentagon):
\begin{equation}
\mathscr{A}_{1IJ}\times\mathscr{A}_{J456}+\mathscr{A}_{1I4J}\times\mathscr{A}_{J56}+\mathscr{A}_{I4 J}\times\mathscr{A}_{1J56}
\end{equation}
and all the vertices in \eqref{12} are given by
\begin{center}
$Z^{(0)}:X_{26}X_{36}X_{46}$,
$Z^{(1)}_{24}:X_{26}X_{24}X_{46}$,
$Z_{1}:X_{26}X_{24}X_{25}$,
$Z_{2}:X_{15}X_{24}X_{25}$,\\
$Z_{3}:X_{14}X_{15}X_{24}$,
$Z_{4}:X_{14}X_{24}X_{46}$
\end{center}
Similar to the $5$-pt case, numerators here have simple structures as they are always determinants of vertices (which are linear combinations of $C_{ij}$'s); for instance the numerator of the first term:
\be
\begin{split}
<Z^{(0)} Z^{(1)}_{24}Z_{1}Z_{2}>=
\Biggm|\begin{matrix}
1 &0 &0 &0\\
1 &0 &C_{24}+C_{25} &0\\
1 &0 &C_{24}+C_{25} &C_{25}+C_{35}\\
1 &C_{15} &C_{15}+C_{24}+C_{25} &C_{15}+C_{25}+C_{35}\\
\end{matrix}\Biggm|
=C_{15} (C_{24}+C_{25}) (C_{25}+C_{35})
\end{split}
\ee
and similarly for the denominators, which give the entire term as:
\be
\begin{split}
&[Z^{(0)},Z^{(1)}_{24},Z_{1},Z_{2}]=\frac{<Z^{(0)} Z^{(1)}_{24}Z_{1}Z_{2}>^3}{<YZ^{(0)} Z^{(1)}_{24}Z_{1}><YZ^{(0)} Z^{(1)}_{24}Z_{2}><YZ^{(0)} Z_{1}Z_{2}><YZ^{(1)}_{24}Z_{1}Z_{2}>}\\
&=\frac{C_{15} (C_{24}+C_{25}) (C_{25}+C_{35})}{X_{24}X_{26}Y_{25}^{24}W_1}
\end{split}
\ee
It is thus straightforward to compute these 3 terms and the sum gives
\be\label{term 24}
\begin{split}
\frac{C_{15} (C_{24}+C_{25}) (C_{25}+C_{35})}{X_{24}X_{26}Y_{25}^{24}W_1}+\frac{(C_{14}+C_{15}) (C_{24}+C_{25}) (C_{15}+C_{25}+C_{35})}{X_{24}X_{46}Y_{14}^{24}W_2}\\
-\frac{C_{14} (C_{24}+C_{25}) (C_{15}+C_{25}+C_{35})^2}{X_{24}Y_{15}^{24}W_1W_2}
\end{split}
\ee
where $W_i$s are new spurious poles introduced in the second step of the recursion:
\begin{equation}
W_1=C_{15} (X_{26}-X_{46})+X_{26} (C_{25}+C_{35})
\end{equation}
and
\begin{equation}
W_2=C_{14} X_{46}-C_{15} (X_{26}-X_{46})-C_{25} X_{26}-C_{35} X_{26}
\end{equation}
Result \eqref{term 24} is identical to the fourth term in the sum \eqref{6}, as can be easily checked. Similar computations give other terms. For instance term $\mathscr{A}_{123I}\times\mathscr{A}_{I456}$ now reads:
\be
\frac{C_{13} (C_{15}+C_{25}+C_{35}) (C_{14}+C_{15}+C_{24}+C_{25})^2}{X_{14}X_{46}Y^{14}_{13}W_3}+\frac{C_{13} (C_{15}+C_{25}+C_{35}) (C_{14}+C_{15}+C_{24}+C_{25})^3}{X_{14}Y^{14}_{24}Y^{14}_{15}W_3}
\ee
where the spurious pole
\be
\begin{split}
W_3=C_{14} (C_{13} X_{46}+C_{15} (X_{36}-X_{26})-C_{25} X_{26}+C_{25} X_{36}-C_{35} X_{26}+C_{35} X_{36})\\
+C_{15} (C_{13} X_{46}-C_{24} X_{26}-2 C_{25} X_{26}+C_{25} X_{36}-C_{35} X_{26}+C_{35} X_{36})\\
-(C_{24}+C_{25}) (-C_{13} X_{46}+C_{25} X_{26}+C_{35} X_{26})+C_{15}^2 (X_{36}-X_{26})
\end{split}
\ee
Note that this is a cubic spurious pole, as opposed to $W_1$, $W_2$ and $Y$'s which are all quadratic. In the sum this spurious pole is canceled, and one gets a result identical to the second term in \eqref{6}.
\paragraph{case $n=7$}
We now move to the next case, $n=7$. The $n=7$ associahedron is a $4-$dim polytope with $14$ $3-$dim facets: $7$ of them are $n=6$ associahedra $\mathscr{A}_{3}$ ($X_{13}$, $X_{16}$, $X_{24}$, $X_{27}$, $X_{35}$, $X_{46}$, $X_{57}$), and $7$ are pentagonal prisms $\mathscr{A}_{2}\times \mathscr{A}_{1}$  ($X_{14}$, $X_{15}$, $X_{25}$, $X_{26}$, $X_{36}$, $X_{37}$, $X_{47}$). This polytope has $42$ vertices. The ``BCFW" representation of the amplitude reads
\begin{equation}
A_{1...7}(X, C)=\sum_{i_{1},j_{1}}\sum_{i_{2},j_{2}}\sum_{a}[Z^{(0)}, Z^{(1)}_{i_1, j_1},Z^{(2)}_{i_2, j_2},Z_a,Z_{a+1}]
\end{equation}
Each term is a canonical function of a $4-$dim simplex. Take the usual basis $\{X_{27},X_{37},X_{47},X_{57}\}$, and the first sum is over $10$ terms. Each of them is again a sum of several "R-invariants". To illustrate the structure in the sum, let's look at one term $\mathscr{A}_{12I67}\times\mathscr{A}_{345I}$, which is the contribution from $\hat X_{36}=0$. This facet $X_{36}$ is of the form:\\
\begin{center}
\begin{tikzpicture}
\draw[black,thick](-1,0,0)--(1,0,0)--(2,0.75,0)--(0,1.25,0)--(-2,0.75,0)--cycle;
\draw[black,thick](-1,0,0)--(1,0,0)--(2.5,0,4)--(0.5,0,4)--cycle;
\draw[black,thick](-2,0.75,0)--(-1,0,0)--(0.5,0,4)--(-1.1,0,2.45)--cycle;
\draw[black,thick](1,0,0)--(2,0.75,0)--(3.5,0.75,4)--(2.5,0,4)--cycle;
\draw[black,dashed](0,1.25,0)--(1.5,1.05,4);
\draw[black,dashed](3.5,0.75,4)--(1.5,1.05,4);
\draw[black,dashed](-1.1,0,2.45)--(1.5,1.05,4);
\filldraw[black] (-1.75,1.5,0) node[anchor=west] {$X_{26}$};
\filldraw[black] (1.75,1.5,0) node[anchor=east] {$X_{16}$};
\filldraw[black] (-2.35,-1.75,0) node[anchor=west] {$X_{27}$};
\filldraw[black](0.6,0,4) node[anchor=north]{$Z^{(1)}_{36}$};
\filldraw[black] (2.25,-1.75,0) node[anchor=east] {$X_{13}$};
\filldraw[black] (0,-1.95,0) node[anchor=north] {$X_{37}$};
\filldraw[black] (-2.5,1.5,0) node[anchor=east] {$upside:X_{46}$};
\filldraw[black] (2.5,-1.5,0) node[anchor=west] {$downside:X_{35}$};
\end{tikzpicture}\\
Facet $X_{36}$ of the $n=7$ associahedron
\end{center}
and we can choose the origin of the second step,  $Z^{(1)}_{36}$, to be the vertex $X_{27}X_{35}X_{36}X_{37}$. Amplitude of this term is then a $9-$term sum:
\be
\begin{split}
[Z^{(0)},Z^{(1)}_{36},Z^{(2)}_{46},Z_1,Z_2]+[Z^{(0)},Z^{(1)}_{36},Z^{(2)}_{46},Z_2,Z_3]+[Z^{(0)},Z^{(1)}_{36},Z^{(2)}_{46},Z_3,Z_4]\\
+[Z^{(0)},Z^{(1)}_{36},Z^{(2)}_{13},Z^{(2)}_{16},Z_2]+[Z^{(0)},Z^{(1)}_{36},Z^{(2)}_{13},Z_2,Z_1]+[Z^{(0)},Z^{(1)}_{36},Z^{(2)}_{16},Z^{(2)}_{26},Z_3]\\
+[Z^{(0)},Z^{(1)}_{36},Z^{(2)}_{16},Z_3,Z_2]+[Z^{(0)},Z^{(1)}_{36},Z^{(2)}_{26},Z_5,Z_4]+[Z^{(0)},Z^{(1)}_{36},Z^{(2)}_{26},Z_4,Z_3]
\end{split}
\ee
where we can easily compute the coordinate of all the vertices appeared:
\begin{center}
$Z^{(0)}:X_{27}X_{37}X_{47}X_{57}, Z^{(2)}_{46}:X_{27}X_{37}X_{36}X_{46}, Z_{1}:X_{13}X_{37}X_{36}X_{46}, Z_{2}:X_{13}X_{16}X_{36}X_{46}$, \\
$Z_{3}:X_{16}X_{26}X_{36}X_{46}, Z_{4}:X_{27}X_{26}X_{36}X_{46}, Z^{(2)}_{13}:X_{13}X_{37}X_{36}X_{35}, Z^{(2)}_{16}:X_{13}X_{16}X_{36}X_{35}$,
$Z^{(2)}_{26}:X_{16}X_{26}X_{36}X_{35}, Z_{5}:X_{27}X_{26}X_{36}X_{35}$
\end{center}
A straightforward but tedious computation gives explicit result of this term, and similarly the entire amplitude $A_{1,2,\cdots, 7}$.
\paragraph{case $n=8$}
With no difficulty one can proceed to higher points, {\it e.g.} the BCFW representation for amplitude with $n=8$ reads:
\begin{equation}
A_{1...8}(X,C)=\sum_{i_{1},j_{1}}\sum_{i_{2},j_{2}}\sum_{i_{3},j_{3}}\sum_{a}[Z^{(0)}, Z^{(1)}_{i_1, j_1},Z^{(2)}_{i_2, j_2},Z^{(3)}_{i_3, j_3},Z_a,Z_{a+1}]
\end{equation}
whose amplituhedron is an $5-$dim associahedron with $20$ $4-$dim facets. 8 of them ($X_{13}$, $X_{17}$, $X_{24}$, $X_{28}$, $X_{35}$, $X_{46}$, $X_{57}$, $X_{68}$) are associahedra $\mathscr{A}_{4}$; 8 $\mathscr{A}_{1}\times \mathscr{A}_{3}$($X_{14}$, $X_{16}$, $X_{25}$, $ X_{27}$, $X_{36}$, $X_{38}$, $X_{47}$, $X_{58}$) ; and rest of them ($X_{15}$, $X_{26}$, $X_{37}$, $X_{48}$) are $\mathscr{A}_{2}\times \mathscr{A}_{2}$. The first sum is then over $15$ terms, as $5$ facets are chosen as basis.\\

\subsection{Properties of the spurious poles}

Before closing, let's briefly discuss the spurious poles one may encounter in various steps of the recursion in (\ref{8}), which are internal facets of a triangulation. We have already seen that besides the spurious poles in the first step, which we have denoted as $Y_{kl}^{ij}$, spurious poles that are introduced in the latter steps are generally different. Of course, now we see that all the poles (physical and spurious) in \eqref{8} are always of the form:
\begin{equation}\label{9}
<YZ_{1}...Z_{n-3}>
\end{equation}
up to a overall factor, where $Y=(1,{\bf X})$ is the vector made up by the basis $X$'s, and all the $n-3$ $Z_{i}$s in (\ref{9}) should be selected from the vertices of a simplex. If $Z^{(0)}$ doesn't appear in (\ref{9}), the pole will be physical. Otherwise, (\ref{9}) is generically a spurious pole, but it can also be a physical one  if the facet spanned by those $Z_{i}$s turns out to be an external facet. It is easy to see that for any $n$, the spurious poles are at most $(n-3)$ power in $X$s and $C$'s as they are derived from $(n-3)$-order determinants.
\paragraph{case $n=6$}
Spurious poles which cannot be represented by $Y_{kl}^{ij}$ first appear when $n=6$. There are not only $\hat X_{ij}(z_{kl})$s (for example the pole $W_2$), but also poles of third power in $X_{ij}$s ($W_3$, for instance). Generally, all the spurious poles in $n=6$ situation are quadratic or cubic: one can easily check that when $Z^{(1)}_{ij}$ is separated from $Z^{(0)}=X_{26}X_{36}X_{46}$ by only one edge, we have only quadratic pole, but otherwise we can have cubic poles.
\paragraph{case $n=7$}
Now we extend the discussion to $n=7$. Without losing generality, we choose the simplex below to illustrate the property.
\begin{center}
\begin{tikzpicture}
\draw[black,thick](-1,0,0)--(1,0,0)--(2,0.75,0)--(0,1.25,0)--(-2,0.75,0)--cycle;
\draw[black,thick](-1,0,0)--(1,0,0)--(2.5,0,4)--(0.5,0,4)--cycle;
\draw[black,thick](-2,0.75,0)--(-1,0,0)--(0.5,0,4)--(-1.1,0,2.45)--cycle;
\draw[black,thick](1,0,0)--(2,0.75,0)--(3.5,0.75,4)--(2.5,0,4)--cycle;
\draw[black,dashed](0,1.25,0)--(1.5,1.05,4);
\draw[black,dashed](3.5,0.75,4)--(1.5,1.05,4);
\draw[black,dashed](-1.1,0,2.45)--(1.5,1.05,4);
\draw[black,dashed](0.5,0,4)--(1,0,0);
\draw[black,dashed](0,1.25,0)--(1,0,0);
\draw[black,dashed](-2,0.75,0)--(1,0,0);
\draw[black,dashed](0.5,0,4)--(-2,0.75,0);
\draw[black,dashed](0.5,0,4)--(0,1.25,0);
\filldraw[black] (-1.75,1.5,0) node[anchor=west] {$X_{26}$};
\filldraw[black] (1.75,1.5,0) node[anchor=east] {$X_{16}$};
\filldraw[black] (-2.35,-1.75,0) node[anchor=west] {$X_{27}$};
\filldraw[black](0.6,0,4) node[anchor=north]{$Z_0$};
\filldraw[black] (2.25,-1.75,0) node[anchor=east] {$X_{13}$};
\filldraw[black] (0,-1.95,0) node[anchor=north] {$X_{37}$};
\filldraw[black] (-2.5,1.5,0) node[anchor=east] {$upside:X_{46}$};
\filldraw[black] (2.5,-1.5,0) node[anchor=west] {$downside:X_{35}$};
\end{tikzpicture}\\
Facet $X_{36}$ of the $n=7$ associahedron and the simplex we choose\\
(with extra vertex $X_{27}X_{37}X_{47}X_{57}$)
\end{center}
where the points appear in the representation of "R-invariants" are chosen as:
\begin{center}
$Z_{\star}:X_{27}X_{37}X_{47}X_{57}$, $Z_0:X_{27}X_{37}X_{35}X_{36}$, $Z_{1}:X_{13}X_{36}X_{37}X_{46}$, \\
$Z_2:X_{26}X_{27}X_{36}X_{46}$, $Z_3:X_{16}X_{26}X_{36}X_{46}$
\end{center}
Thereby the poles produced by "R-invariant" $[Z_{\star},Z_0,Z_1,Z_2,Z_3]$ finally read:
\begin{equation}
\begin{split}
U_1=(C_{35}+C_{36}) (-(C_{13}+C_{14}+C_{15}+C_{16}) (X_{37} (C_{16}+C_{26}+C_{36}+C_{46})-X_{57} (C_{16}+C_{26}))\\
-(C_{36}+C_{46}) (C_{16} X_{27}-C_{16} X_{37}+C_{26} X_{27}))\phantom{aaaaaaaaaaaaaa}\\
-(C_{36}+C_{46}) (-(C_{13}+C_{14}+C_{15}+C_{16}) (C_{16} X_{37}-C_{16} X_{47}+C_{26} X_{37}-C_{26} X_{47}+C_{36} X_{37})\\
-C_{36} (C_{16} X_{27}-C_{16} X_{37}+C_{26} X_{27}))\phantom{aaaaaaaaaaa}
\end{split}
\end{equation}
\begin{equation}
\begin{split}
U_2=(C_{35}+C_{36}) (-C_{26} (X_{27} (C_{36}+C_{46})-X_{57} (C_{13}+C_{14}+C_{15}+C_{16}))\\
-X_{37} (C_{26}+C_{36}+C_{46}) (C_{13}+C_{14}+C_{15}+C_{16}))\\
-(C_{36}+C_{46}) (X_{37} (C_{26}+C_{36}) (-(C_{13}+C_{14}+C_{15}+C_{16}))\\
-C_{26} (C_{36} X_{27}-X_{47} (C_{13}+C_{14}+C_{15}+C_{16})))
\end{split}
\end{equation}
two quartic poles and:
\begin{equation}
U_3=C_{26} (C_{35} (X_{57}-X_{37})+C_{36} X_{57}-C_{36} X_{47}+C_{46} X_{37}-C_{46} X_{47})+C_{35} (C_{36}+C_{46}) (X_{27}-X_{37})
\end{equation}
\begin{equation}
U_4=C_{36} (X_{57}-X_{47})+C_{46} (X_{37}-X_{47})
\end{equation}
a cubic pole and a square pole. All these spurious poles now are up to the forth power in $X_{ij}$s, also all the spurious poles produced by the recursion when $n=7$. Those quartic poles only appear when coordinates for two vertices differ in at least two components, otherwise they only contribute an overall factor when taking the determinant, which is canceled by the numerator and gives lower-order poles.

\section{Discussions}\label{sec 4}

In this note we have derived a recursion relation, which is similar to the BCFW one, for tree-level amplitudes in bi-adjoint $\phi^3$ theory, which directly corresponds to triangulations of the kinematic associahedron underlying the amplitude. The key ingredient is a one-parameter deformation in $n{-}3$ independent kinematic variables, $X_{i,j} \to z X_{i,j}$ while keeping all constants defining the subspace, $C_{i,j}$, undeformed. With a suitable prefactor such that there is no pole at $z=0$ or $z=\infty$, the amplitude is then a sum of residues at finite $z$ which factorize into lower-point amplitudes/associahedra. This gives a partial triangulation into $(n{-}2)(n{-}3)/2$ polytopes by connecting the origin to all remaining facets, each corresponding to a factorization channel. By applying the recursion $n{-}3$ times, we have a solution which gives a full triangulation of the associahedron into simplices; each term is given by the canonical function of a simplex, in analogy with the ``R invariant" of ${\cal N}=4$ SYM amplitude. Different ways for choosing the origin in each step give all triangulations of the associahedron, including ``inside" and ``outside" ones, and the general formula gives explicit results for all of them.

These ``BCFW" representations of the $\phi^3$ amplitude make certain properties of the amplitude manifest, such as projectivity and ``soft limit", which are obscured in Feynman diagrams. Recall that each BCFW term/cell of amplituhedron in ${\cal N}=4$ SYM enjoys the Yangian symmetry({\it c.f.}~\cite{Grassmannian}), and it is highly desirable to see if other non-trivial properties or symmetries can be made manifest in our representation. We have focused on $m(\alpha|\beta)$ with $\alpha=\beta$, but since our recursion relation applies to the canonical form, $\Omega(\mathscr{A}_{n{-}3})$, we can get similar representation for general $m(\alpha|\beta)$ by {\it e.g.} pullback to the subspace with a different ordering. Geometrically they correspond to degenerate cases where some vertices are taken to infinity, whose triangulations can be worked out similarly. It would be interesting to relate our construction to that of~\cite{Frost}. Furthermore, an important open question is to study recursion relations for loop integrands of the theory, in particular at one-loop they should be  related to triangulations of the halohedron in~\cite{Halohedron}.

One can also consider recursion relations beyond amplitudes in bi-adjoint $\phi^3$ theory. For example, can we find similar recursion relations for scalar amplitudes with quartic or higher-order vertices? More importantly, can we apply our recursion to amplitudes in general massless theories, such as Yang-Mills and non-linear sigma model? If we naively apply the one-parameter deformation, generically there is a pole at infinity since the kinematic numerators now also depend on $z$. However, if we distinguish $k\cdot k$ in the numerators from the poles $X_{i,j}$ and only deform the latter (see discussions in~\cite{Arkani-Hamed:2017thz}), we immediately obtain recursion relations for these theories as well. It is not clear to us yet in practice how useful would such recursion relations be, and we leave these issues for future investigations. It would also be very interesting to compare our recursion for $\phi^3$ theory with other recursion relations for scalar theories, such as that for special effective field theories~\cite{EFT}, which may hint at possible geometric structures underlying those amplitudes.

Last but not least, let's comment on some results that have not been discussed in the main text. As is familiar from the amplituhedron story, while BCFW representations can be obtained from triangulating the associahedron, one gets local representations by considering triangulations of the dual; for completeness, we collect explicit formulas for them in appendix A.  Moreover, essentially the same recursion relations can be derived for the canonical form/function of Cayley polytopes in kinematic space~\cite{Gao:2017iop}, which are also sum of cubic trees. We present such recursions that also give triangulations of Cayley polytopes (with an example for permutohedron) in appendix C. It would be interesting to further explore triangulations of more general polytopes, and possible relations with Jeffrey-Kirwan residues~\cite{JK} and generalized associahedra~\cite{2018arXiv180809986B}.
\section*{Acknowledgments}

S.H. thanks Nima Arkani-Hamed for suggesting the problem, and also Yuntao Bai and Gongwang Yan for collaborating on related projects. We thank Chi Zhang,  Yong Zhang for stimulating discussions. SH's research is supported in part by the Thousand Young Talents program, the Key Research Program of Frontier Sciences of CAS under Grant No. QYZDB-SSW-SYS014 and Peng Huanwu center under Grant No. 11747601.

\appendix
\section{Coordinates and triangulations}
In this appendix we collect results for the coordinate for vertices of the associahedron and those of the dual associahedron. As we use the former for triangulation in the main text, here we use the former to give local representation of the amplitude from triangulation of the dual associahedron.
\subsection*{Coordinate for cubic trees from mutations}
We have already explained how to obtain the coordinate for any vertex, or cubic tree, by solving linear equations. It turns out that the result has a very interesting pattern, which allows one to directly read it off by considering how the cubic tree is obtained from an initial one via a series of mutations. We use a $5-$point example to illustrate this procedure.\\
Firstly, we choose $X_{25}-X_{35}$ as the basis, then the cubic tree:\\
\begin{center}
\begin{tikzpicture}
\draw[gray,thick](0,0)--(0,-1);
\draw[gray,thick](0,0)--(1.5,1.5);
\draw[gray,thick](0,0)--(-1.5,1.5);
\draw[gray,thick](0.5,0.5)--(-0.5,1.5);
\draw[gray,thick](1,1)--(0.5,1.5);
\filldraw[black] (0,-1) node[anchor=north] {5};
\filldraw[black] (1.5,1.5) node[anchor=south] {4};
\filldraw[black] (-1.5,1.5) node[anchor=south] {1};
\filldraw[black] (-0.5,1.5) node[anchor=south] {2(x)};
\filldraw[black] (0.5,1.5) node[anchor=south] {3(y)};
\end{tikzpicture}
\end{center}
will be the origin. Note that when draw the cubic trees, we will always put the $5$th-particle as the stem and the $1$st and $4$th-particles as the first and the last leaf. So that only the two leaves between them could move. Similarly the $n-$particles case, $n-3$ of whose leaves could be muted. We assign coordinates $(x,y)$ (or $(x_{1},...,x_{n-3})$ for $n-$case) respectively to the $2$nd and $3$rd leaves.\\
Now we define a mutation rule of the tree and respectively the coordinate as:

\begin{itemize}
\item[1)]
Every bare leaf could be muted between the crotch held by the nearest lower node. After the mutation, the new node the leaf found should also be next to the lower node.
\item[2)]
When one mutation is done, the corresponding component of the coordinates will plus (right to left) or minus (left to right) a linear combination of $C_{ij}$s, whose first index $i$ are selected from the LHS indexes and the second index $j$ from the RHS.
\end{itemize}

This rule defines all the coordinates. The figure below is the example when $n=5$:\\
\begin{center}
\begin{tikzpicture}
\draw[gray,thick](0,0)--(0,-1);
\draw[gray,thick](0,0)--(1.5,1.5);
\draw[gray,thick](0,0)--(-1.5,1.5);
\draw[gray,thick](0.5,0.5)--(-0.5,1.5);
\draw[gray,thick](0,1)--(0.5,1.5);
\filldraw[black] (0,-1) node[anchor=north] {5};
\filldraw[black] (1.5,1.5) node[anchor=south] {4};
\filldraw[black] (-1.5,1.5) node[anchor=south] {1};
\filldraw[black] (-0.5,1.5) node[anchor=south] {2(x)};
\filldraw[black] (0.5,1.5) node[anchor=south] {3(y)};
\filldraw[black] (1,-0.5) node[anchor=north] {(0,$C_{24}$)};
\end{tikzpicture}
$\leftarrow$
\begin{tikzpicture}
\draw[gray,thick](0,0)--(0,-1);
\draw[gray,thick](0,0)--(1.5,1.5);
\draw[gray,thick](0,0)--(-1.5,1.5);
\draw[gray,thick](0.5,0.5)--(-0.5,1.5);
\draw[gray,thick](1,1)--(0.5,1.5);
\filldraw[black] (0,-1) node[anchor=north] {5};
\filldraw[black] (1.5,1.5) node[anchor=south] {4};
\filldraw[black] (-1.5,1.5) node[anchor=south] {1};
\filldraw[black] (-0.5,1.5) node[anchor=south] {2(x)};
\filldraw[black] (0.5,1.5) node[anchor=south] {3(y)};
\filldraw[black] (1,-0.5) node[anchor=north] {(0,0)};
\end{tikzpicture}
$\rightarrow$
\begin{tikzpicture}
\draw[gray,thick](0,0)--(0,-1);
\draw[gray,thick](0,0)--(1.5,1.5);
\draw[gray,thick](0,0)--(-1.5,1.5);
\draw[gray,thick](-1,1)--(-0.5,1.5);
\draw[gray,thick](1,1)--(0.5,1.5);
\filldraw[black] (0,-1) node[anchor=north] {5};
\filldraw[black] (1.5,1.5) node[anchor=south] {4};
\filldraw[black] (-1.5,1.5) node[anchor=south] {1};
\filldraw[black] (-0.5,1.5) node[anchor=south] {2(x)};
\filldraw[black] (0.5,1.5) node[anchor=south] {3(y)};
\filldraw[black] (1.25,-0.5) node[anchor=north] {($C_{13}+C_{14}$,0)};
\end{tikzpicture}
$\rightarrow$
\begin{tikzpicture}
\draw[gray,thick](0,0)--(0,-1);
\draw[gray,thick](0,0)--(1.5,1.5);
\draw[gray,thick](0,0)--(-1.5,1.5);
\draw[gray,thick](-1,1)--(-0.5,1.5);
\draw[gray,thick](-0.5,0.5)--(0.5,1.5);
\filldraw[black] (0,-1) node[anchor=north] {5};
\filldraw[black] (1.5,1.5) node[anchor=south] {4};
\filldraw[black] (-1.5,1.5) node[anchor=south] {1};
\filldraw[black] (-0.5,1.5) node[anchor=south] {2(x)};
\filldraw[black] (0.5,1.5) node[anchor=south] {3(y)};
\filldraw[black] (2,-0.5) node[anchor=north] {($C_{13}+C_{14}$,$C_{14}+C_{24}$)};
\end{tikzpicture}
$\rightarrow$
\begin{tikzpicture}
\draw[gray,thick](0,0)--(0,-1);
\draw[gray,thick](0,0)--(1.5,1.5);
\draw[gray,thick](0,0)--(-1.5,1.5);
\draw[gray,thick](0,1)--(-0.5,1.5);
\draw[gray,thick](-0.5,0.5)--(0.5,1.5);
\filldraw[black] (0,-1) node[anchor=north] {5};
\filldraw[black] (1.5,1.5) node[anchor=south] {4};
\filldraw[black] (-1.5,1.5) node[anchor=south] {1};
\filldraw[black] (-0.5,1.5) node[anchor=south] {2(x)};
\filldraw[black] (0.5,1.5) node[anchor=south] {3(y)};
\filldraw[black] (1.5,-0.5) node[anchor=north] {($C_{14}$,$C_{14}+C_{24}$)};
\end{tikzpicture}
\end{center}
A similar operation gives us the coordinates of the $6-$points cubic trees. For example, if we want to obtain the coordinates of the tree (vertex $X_{14}X_{24}X_{46}$ of the associahedron):
\begin{center}
\begin{tikzpicture}
\draw[gray,thick](0,0)--(0,-1);
\draw[gray,thick](0,0)--(2,2);
\draw[gray,thick](0,0)--(-2,2);
\draw[gray,thick](-1,1)--(0,2);
\draw[gray,thick](-0.5,1.5)--(-1,2);
\draw[gray,thick](1.5,1.5)--(1,2);
\filldraw[black](0,-1)node[anchor=north] {6};
\filldraw[black] (2,2) node[anchor=south] {5};
\filldraw[black] (-2,2) node[anchor=south] {1};
\filldraw[black] (-1,2) node[anchor=south] {2(x)};
\filldraw[black] (0,2) node[anchor=south] {3(y)};
\filldraw[black] (1,2) node[anchor=south] {4(z)};
\end{tikzpicture}
\end{center}
We do the following mutation:
\begin{center}
\begin{tikzpicture}
\draw[gray,thick](0,0)--(0,-1);
\draw[gray,thick](0,0)--(2,2);
\draw[gray,thick](0,0)--(-2,2);
\draw[gray,thick](1,1)--(0,2);
\draw[gray,thick](0.5,0.5)--(-1,2);
\draw[gray,thick](1.5,1.5)--(1,2);
\filldraw[black] (0,-1)node[anchor=north] {6};
\filldraw[black] (2,2) node[anchor=south] {5};
\filldraw[black] (-2,2) node[anchor=south] {1};
\filldraw[black] (-1,2) node[anchor=south] {2(x)};
\filldraw[black] (0,2) node[anchor=south] {3(y)};
\filldraw[black] (1,2) node[anchor=south] {4(z)};
\filldraw[black] (2,-1) node[anchor=north] {(0,0,0)};
\end{tikzpicture}
$\rightarrow$
\begin{tikzpicture}
\draw[gray,thick](0,0)--(0,-1);
\draw[gray,thick](0,0)--(2,2);
\draw[gray,thick](0,0)--(-2,2);
\draw[gray,thick](1,1)--(0,2);
\draw[gray,thick](-1.5,1.5)--(-1,2);
\draw[gray,thick](1.5,1.5)--(1,2);
\filldraw[black](0,-1)node[anchor=north] {6};
\filldraw[black] (2,2) node[anchor=south] {5};
\filldraw[black] (-2,2) node[anchor=south] {1};
\filldraw[black] (-1,2) node[anchor=south] {2(x)};
\filldraw[black] (0,2) node[anchor=south] {3(y)};
\filldraw[black] (1,2) node[anchor=south] {4(z)};
\filldraw[black] (2,-1) node[anchor=north] {($C_{13}+C_{14}+C_{15}$,0,0)};
\end{tikzpicture}
$\rightarrow$
\begin{tikzpicture}
\draw[gray,thick](0,0)--(0,-1);
\draw[gray,thick](0,0)--(2,2);
\draw[gray,thick](0,0)--(-2,2);
\draw[gray,thick](-1,1)--(0,2);
\draw[gray,thick](-1.5,1.5)--(-1,2);
\draw[gray,thick](1.5,1.5)--(1,2);
\filldraw[black](0,-1)node[anchor=north] {6};
\filldraw[black] (2,2) node[anchor=south] {5};
\filldraw[black] (-2,2) node[anchor=south] {1};
\filldraw[black] (-1,2) node[anchor=south] {2(x)};
\filldraw[black] (0,2) node[anchor=south] {3(y)};
\filldraw[black] (1,2) node[anchor=south] {4(z)};
\filldraw[black] (1,-2) node[anchor=north] {($C_{13}+C_{14}+C_{15}$,$C_{14}+C_{15}+C_{24}+C_{25}$,0)};
\end{tikzpicture}
$\rightarrow$
\begin{tikzpicture}
\draw[gray,thick](0,0)--(0,-1);
\draw[gray,thick](0,0)--(2,2);
\draw[gray,thick](0,0)--(-2,2);
\draw[gray,thick](-1,1)--(0,2);
\draw[gray,thick](-0.5,1.5)--(-1,2);
\draw[gray,thick](1.5,1.5)--(1,2);
\filldraw[black](0,-1)node[anchor=north] {6};
\filldraw[black] (2,2) node[anchor=south] {5};
\filldraw[black] (-2,2) node[anchor=south] {1};
\filldraw[black] (-1,2) node[anchor=south] {2(x)};
\filldraw[black] (0,2) node[anchor=south] {3(y)};
\filldraw[black] (1,2) node[anchor=south] {4(z)};
\filldraw[black] (1,-2) node[anchor=north] {($C_{14}+C_{15}$,$C_{14}+C_{15}+C_{24}+C_{25}$,0)};
\end{tikzpicture}
\end{center}
\subsection*{Triangulation of the dual and local representation}
Here we present local representations of the amplitude by triangulating the dual associahedron. Note that this has been done in \cite{Arkani-Hamed:2017thz}, but we work these out more explicitly here for completeness. Since the dual associahedron is a simplicial polytope, its triangulation can be trivially done by connecting a vertex (or a facet of the associahedron) to all other vertices. Recall that once a basis is chosen, our basic formula  $X_{a,b}=X^0_{a,b}+c_{a,b}$ can be equivalently written as
\begin{equation}
X_{a,b}=Y\cdot W_{a,b}
\end{equation}
where $Y=(1, \bf{X})$ and $W_{a,b}$ is nothing but the coordinate of this vertex in the dual associahedron. For example, for $n=6$ with $Y=(1,X_{26},X_{36},X_{46})$, we can easily work out all the 9 $W_{a,b}$'s, and we will use the following two shortly:
\begin{equation}
\begin{split}
&W_{13}=(C_{13}+C_{14}+C_{15},-1,0,0)\,,\quad W_{25}=(C_{25}+C_{35},1,0,-1)
\end{split}
\end{equation}
The amplitude $A_{1,2,\cdots,n}$ can be written as the volume of the dual associahedron, $\mathscr{A^\star_n}$; the latter can be written as a sum of volume of simplices, which are formed by a reference point $W_0$ and all facets (or vertices of the associahedron) $Z$:
\begin{equation}
Vol(\mathscr{A^\star_n})=\frac1{Y\cdot W_{0}}\sum_Z \frac{Z\cdot W_{0}}{\prod_{a=1}^{n-3} Y\cdot W_{i_a j_a}}
\end{equation}
where we denote the $n{-}3$ vertices of the facet $Z$ as $W_{i_a, j_a}$ for $a=1,\cdots, n{-}3$; all the poles are of the form $Y \cdot W_{i,j}=X_{i,j}$, which is why this gives the local representation for the amplitude.\\
For instance, the result of $n=5$ situation, applying the coordinates of the vertices, is immediately:
\begin{equation}
Vol(\mathscr{A^\star_5})=\frac1{X_{13}}(\frac{C_{13}}{X_{14}X_{24}}+\frac{C_{13}+C_{14}}{X_{25}X_{24}}+\frac{C_{13}+C_{14}}{X_{25}X_{35}})
\end{equation}
Things go a little bit more complicated when number of particles rises: the number of terms may vary between different references. \\
Let't turn to the $6-$points examples. Firstly, choose $W_{13}$ as a reference:
\begin{equation}
\begin{split}
&Vol(\mathscr{A^\star_6})=\frac1{X_{13}}(\frac{C_{13}}{X_{14}X_{15}X_{24}}+\frac{C_{13}+C_{14}}{X_{15}X_{25}X_{24}}+\frac{C_{13}}{X_{14}X_{46}X_{24}}+\frac{C_{13}+C_{14}+C_{15}}{X_{24}X_{25}X_{26}}+\\
&\frac{C_{13}+C_{14}+C_{15}}{X_{46}X_{26}X_{24}}+\frac{C_{13}+C_{14}+C_{15}}{X_{25}X_{35}X_{26}}+\frac{C_{13}+C_{14}+C_{15}}{X_{26}X_{35}X_{36}}+\frac{C_{13}+C_{14}+C_{15}}{X_{26}X_{36}X_{46}}+\frac{C_{13}+C_{14}}{X_{15}X_{25}X_{35}})
\end{split}
\end{equation}
Changing to another reference, for instance $X_{25}$, the result reads:
\begin{equation}
\begin{split}
Vol(\mathscr{A^\star_6})=\frac1{X_{25}}(\frac{C_{13}+C_{14}}{X_{13}X_{14}X_{15}}+\frac{C_{13}+C_{14}}{X_{13}X_{35}X_{15}}+\frac{C_{13}+C_{14}+C_{15}+C_{25}}{X_{13}X_{35}X_{36}}+\frac{C_{14}}{X_{24}X_{14}X_{15}}\phantom{aaaaaa}\\
+\frac{C_{13}+C_{14}+C_{15}+C_{25}+C_{35}}{X_{13}X_{36}X_{46}}+\frac{C_{13}+C_{14}+C_{15}+C_{25}+C_{35}}{X_{13}X_{14}X_{46}}+\frac{C_{14}+C_{15}+C_{25}+C_{35}}{X_{14}X_{24}X_{46}}\\
+\frac{C_{25}}{X_{35}X_{36}X_{26}}+\frac{C_{25}+C_{35}}{X_{26}X_{36}X_{46}}+\frac{C_{25}+C_{35}}{X_{24}X_{26}X_{46}})
\end{split}
\end{equation}
It could be spotted that the number of the terms has varied as edges of the facets also vary in number. All these results however are non-trivially equal.

\section{Explicit results for $n=6$ amplitude from recursion }\label{16}
We list the numerators of every terms in the first step recursion for $A_{1,2,\cdots,n}$:
\begin{equation}
\begin{split}
N_{13}=-(X_{13}+X_{26}) (-X_{36} (X_{15} (X_{13}^2 (-X_{46})+X_{13} X_{26} (X_{36}-2 X_{46})+X_{26}^2 X_{35})\phantom{aaaaaaaaaa}\\
+X_{46} (X_{13}^2 (X_{35}-X_{36})+X_{13} X_{26} (X_{35}-X_{46})+X_{26}^2 X_{35}))\phantom{aaaaaa}\\
-X_{14} X_{26} (X_{15} (X_{13} X_{36}+X_{26} (X_{35}+X_{46}))+X_{46} (X_{26}X_{36}-X_{13} (X_{35}-2 X_{36}+X_{46}))))
\end{split}
\end{equation}
\begin{equation}
\begin{split}
N_{14}=-X_{36} (X_{14}+X_{36}) (X_{15}+X_{46}) (X_{13}-X_{14}+X_{24}) (2 X_{14}-2 X_{15}-X_{24}+2 X_{25}-X_{26}+X_{36})
\end{split}
\end{equation}
\begin{equation}
\begin{split}
N_{15}=-(X_{15}+X_{46})^3 (X_{46} (X_{13}-X_{14}+X_{24}) (X_{24}-X_{25}+X_{35}) (X_{15} (X_{26}-X_{36})+X_{24} X_{46})\\
+(X_{14}-X_{15}-X_{24}+X_{25}) (X_{15} X_{36}-X_{46} (X_{13}+X_{24})) (X_{15} (X_{46}-X_{26})-X_{46} (X_{24}+X_{35})))
\end{split}
\end{equation}
\begin{equation}
\begin{split}
N_{24}=(X_{24}-X_{26}+X_{36}) ((X_{26}-X_{36}) (X_{25}-X_{26}+X_{46}) (X_{14}-X_{15}-X_{24}+X_{25})\\
 (X_{24} (X_{26}-X_{46})+X_{25} (X_{36}-X_{26}))
+(X_{15}-X_{25}+X_{26}) (X_{24} X_{26}-(X_{25}+X_{46}) (X_{26}-X_{36})) \\
(X_{14} (X_{26}-X_{36})+X_{24} (-X_{26}+X_{36}+X_{46})+X_{25} (X_{26}-X_{36})))
\end{split}
\end{equation}
\begin{equation}
\begin{split}
N_{25}=(X_{26}-X_{46}) (X_{15}-X_{25}+X_{26}) (-X_{24}+X_{25}-X_{35}) (X_{25}-X_{26}+X_{46})^2
\end{split}
\end{equation}
\begin{equation}
\begin{split}
N_{35}=-(X_{35}-X_{36}+X_{46})(-(X_{36}-X_{46}) (X_{13}-X_{15}+X_{25}) (X_{35} X_{46} (X_{25}-X_{35}+X_{36})\\
+X_{15} ((X_{36}-X_{46}) (X_{25}-X_{35}+X_{36})-X_{26} X_{35})+X_{26}(X_{35}^2-2 X_{35} X_{36}+X_{36}^2-X_{36} X_{46}))\\
-(X_{15}-X_{25}+X_{26}) (X_{15} (X_{36}-X_{46})+X_{35} X_{46}) ((X_{36}-X_{46}) (X_{25}-X_{35}+X_{36})-X_{26} X_{35}))
\end{split}
\end{equation}
\section{Cayley polytopes}
Finally, we extend our discussion to general Cayley polytopes~\cite{Gao:2017iop}.
\subsection*{Construction and canonical function of the polytopes}
Cayley polytopes are natural generalization of associahedron which can be constructed from a labelled tree with nodes $\{1,2,...,n-1\}$. In particular, while the associahedron is a Cayley polytope from a linear tree (or Hamiltonian graph), the permutohedron is a Cayley polytope constructed from a star-shaped tree. In kinematic space, any such Cayley polytope can be obtained as the intersection of a positive region and a $n{-}3$-dim subspace very similar to the associahedron. The canonical form/function is again given by sum over cubic trees, though more general than just the planar ones.  For example, an $n=5$ permutohedron constructed from the tree:
\begin{center}
\begin{tikzpicture}
\draw[black,thick](-1,0)--(0,0)--(1,0);
\draw[black,thick](0,0)--(0,1);
\filldraw[black](-1,0) node[anchor=north east]{$1$};
\filldraw[black](0,0) node[anchor=north]{$2$};
\filldraw[black](1,0) node[anchor=north west]{$3$};
\filldraw[black](0,1) node[anchor=south]{$4$};
\end{tikzpicture}
\end{center}
can be defined as the intersection of a region with
\begin{equation}
s_{12},s_{23},s_{24},s_{123},s_{124},s_{234}\geq0
\end{equation}
and a two-dimensional subspace defined by
\begin{equation}
s_{124}=s_{12}+s_{24}-c_{14}
\end{equation}
\begin{equation}
s_{234}=s_{23}+s_{24}-c_{34}
\end{equation}
\begin{equation}
s_{123}=s_{12}+s_{23}-c_{13}
\end{equation}
\begin{equation}
s_{123}+s_{124}+s_{234}=s_{12}+s_{23}+s_{24}
\end{equation}
So that the $n=5$ permutohedron is a hexagon in $2-$dim. Without losing generality, like what we did in the situations of associahedron, one can choose a compatible pair of these Mandelstams, for example $s_{12}$ and $s_{123}$, as basis to express the others. The coordinates for all the vertices of this polygon could then be gained after solving $6$ pairs of linear equations, which finally leads to a polygon:
\begin{center}
\begin{tikzpicture}
\draw[black,thick](0,0)--(0,1.5);
\draw[black,thick](0,1.5)--(1.5,3);
\draw[black,thick](1.5,3)--(3,3);
\draw[black,thick](3,3)--(3,1.5);
\draw[black,thick](3,1.5)--(1.5,0);
\draw[black,thick](1.5,0)--(0,0);
\draw[black,dashed](0,0)--(1.5,3);
\draw[black,dashed](0,0)--(3,3);
\draw[black,dashed](0,0)--(3,1.5);
\filldraw[black](0,0)node[anchor=north east]{$(0,0)$};
\filldraw[black](0,1.5)node[anchor=south east]{$(0,c_{34})$};
\filldraw[black](1.5,3)node[anchor=south east]{$ (c_{14},c_{14}+c_{34})$};
\filldraw[black](3,3)node[anchor=south west]{$(c_{13}+c_{14},c_{14}+c_{34})$};
\filldraw[black](3,1.5)node[anchor=north west]{$(c_{13}+c_{14},c_{14})$};
\filldraw[black](1.5,0)node[anchor=north west]{$(c_{13},0)$};
\filldraw[black](0,0.75)node[anchor=east]{$s_{12}$};
\filldraw[black](0.75,2.25)node[anchor=south east]{$s_{124}$};
\filldraw[black](2.25,3)node[anchor=south]{$s_{24}$};
\filldraw[black](3,2.25)node[anchor=west]{$s_{234}$};
\filldraw[black](2.25,0.75)node[anchor=north west]{$s_{23}$};
\filldraw[black](0.75,0)node[anchor=north]{$s_{123}$};
\end{tikzpicture}
\\the coordinates for vertices and the triangulation of $n=5$ permutohedron
\end{center}
After we obtain the coordinates of the vertices, a direct "inside" triangulation could also be applied to the hexagon to derive its canonical function:
\begin{equation}
\begin{split}
A_{5}=\frac{c_{14} c_{34}}{s_{12}s_{124}(c_{14} (s_{12}-s_{123})+c_{34}s_{12})}+\frac{c_{13} (c_{14}+c_{34})^2}{s_{24}(c_{14} (s_{123}-s_{12})-c_{34} s_{12}) (c_{13} s_{123}+c_{14} (s_{123}-s_{12})-c_{34} s_{12})}\\
+\frac{c_{34} (c_{13}+c_{14})^2}{s_{234} (c_{13} s_{123}+c_{14} (s_{123}-s_{12}))(c_{13} s_{123}+c_{14} (s_{123}-s_{12})-c_{34} s_{12})}+\frac{c_{13} c_{14}}{s_{123} s_{23} (c_{13} s_{123}+c_{14} (s_{123}-s_{12}))}
\end{split}
\end{equation}
\subsection*{Recursion relation for the canonical functions}
Like what happens in associahedra case, the canonical functions of Cayley polytopes also factorize at certain physical poles $s=0$, which are in one-to-one correspondence with the subgraphs of the tree the whole polytope constructed from. At each pole $s_I$, the tree correspondingly factorizes into two pieces: the subgraph corresponds to $s_I$ itself and the rest part of the whole tree when this subgraph degenerate to a point. Rescaling those $s_{I}$ chosen as basis to $z\cdot s_{I}$ similarly, the function could then be expressed by a sum over the canonical functions of smaller trees as:
\begin{equation}\label{cay}
A_{n}=\sum_{\substack{subgraphs\ s_{I}\\except\ basis}}\frac{z_{I}^{n-3}}{s_{I}}A_{s_I}\times A_{R_{I}}
\end{equation}
where the sum is over all the Mandelstam variables appear in the tree except those basis, and $z_{I}$ stands for the solve of $\hat s_{I}(z)=0$. Canonical functions of the factorized trees are noted as $A_{s_I}$ and $A_{R_I}$ on the RHS. They are functions of $z_I$. It is easy to see that the relation \eqref{2} is just a special case of \eqref{cay}, and \eqref{cay} also gives a full triangulation of the Cayley polytope.\\
Take one term of $n=5$ permutohedron, such as $\hat s_{124}=0$, as an example. The Cayley tree accordingly breaks into two parts, an $\mathscr{A}_1$ and an $\mathscr{A}_{0}$ (trivial). With the result of $\mathscr{A}_1$, the term then reads:
\begin{equation}
\frac{z_{124}^2}{s_{124}}(\frac1{\hat s_{12}}+\frac1{\hat s_{24}})=\frac{c_{14} c_{34}}{s_{12}s_{124}(c_{14} (s_{12}-s_{123})+c_{34}s_{12})}
\end{equation}
which is the first term in the canonical function result. Other terms can be obtained likewise.

\bibliographystyle{utphys}
\bibliography{refs}
\end{document}